\documentclass[pra,twocolumn]{revtex4}
\usepackage{graphicx} % standard LaTeX graphics tool
\usepackage{amsmath}
\begin{document}
\title{Quantum teleportation of light beams}
\author{T. C. Zhang}
\author{K. W. Goh}
\author{C. W. Chou}
\author{P. Lodahl}
\author{H. J. Kimble}
\affiliation{Norman Bridge Laboratory of Physics 12-33, California Institute
of Technology, Pasadena, CA 91125}
\date{July 11, 2002}

\begin{abstract}
We experimentally demonstrate quantum teleportation for continuous variables
using squeezed-state entanglement. The teleportation fidelity for a real
experimental system is calculated explicitly, including relevant imperfection
factors such as propagation losses, detection inefficiencies and phase
fluctuations. The inferred fidelity for input coherent states is
$F=0.61\pm 0.02$, which when corrected for the efficiency of detection by
the output observer, gives a fidelity of $0.62$.
By contrast, the projected result based on the independently measured
entanglement and efficiencies is $0.69$. The teleportation protocol is
explained in detail, including a discussion of discrepancy between experiment
and theory, as well as of the limitations of the current apparatus.
\end{abstract}


\maketitle

\section{Introduction}

The \textit{No Cloning Theorem} prohibits making an exact copy of an
unknown quantum state \cite{Wootters82}. Yet, it is nevertheless possible to
transport an unknown quantum state from one place to another without having
the associated physical object propagate through the intervening space by way
of a process termed \textit{quantum teleportation} in the landmark work by
Bennett et al.\ \cite{Bennett93} in 1993. This ``disembodied'' transport
of quantum states is made possible by utilizing
shared quantum entanglement and classical communication between the
sending and receiving locations. In recent years, quantum teleportation has
played a central role in quantum information science and has become an
essential tool in diverse quantum algorithms and
protocols \cite{Preskill98,NielsenChuang00}.

By contrast, progress on an experimental front has been rather more modest
in the actual attainment of quantum teleportation
\cite{Bouwmeester97,Boschi98,Furusawa98,Nielsen98}. An overview of
these various experiments as well as operational criteria for gauging
laboratory success can be found in Refs.\ \cite{Braunstein00,Braunstein01}.
Significantly, to date only the experiment of Furusawa et
al.\ \cite{Furusawa98} on continuous variables has achieved unconditional
quantum teleportation \cite{ANU02}.

The purpose of this paper is to present a report of our progress in the
continuation of the experiment as reported by Furusawa et al.\
\cite{Furusawa98} and as described in Ref.\ \cite{Sorensen98}. We give a
detailed description of our quantum teleportation apparatus and procedures,
and include recent experimental results \cite{osa01}.
Some notable distinctions between
our current experiment and the previous one by Furusawa et al.\ are improved
EPR entanglement, better detection efficiencies, and ultimately, a higher
fidelity between the input and teleported output states. We also investigate
in some detail the various factors that limit the quality of the teleportation
procedure under realistic conditions and as are applicable to the experimental
setup. We provide a detailed model of the entire experiment that includes
essentially all of the dominant loss mechanisms and utilize this model to gain
insight into the limitations of the current apparatus and protocols, and
thereby to discover methods of circumventing these limitations.

Our experiment is based on the continuous variable teleportation protocol
first proposed in \cite{B&K98}, which in turn was motivated by the work of
Vaidman \cite{Vaidman94}. In our realization of this protocol, an entangled
EPR state \cite{EPR} is created from two independent squeezed fields.
One half of this entangled state (called EPR1) is sent to Alice, who in turn
combines it at a 50/50 beamsplitter with an unknown input state that is
intended for teleportation. Note that the input quantum state is unknown to
both Alice and Bob. Alice subsequently measures the $x$ and $p$ quadratures of
the two output fields from the beamsplitter, the $x$ quadrature for one beam
and the $p$ quadrature for the other. This measurement of $(x,p)$ provides the
continuous variable analogy to a Bell-state measurement for the discrete
variable case \cite{vanenk99}. In the limit of perfect EPR
correlations, Alice gains no information about the input state. The output
photocurrents from Alice's two quadrature measurements are transmitted to Bob
via classical information channels. Bob then uses them to perform a continuous
phase space displacement on the second EPR beam (EPR2), thereby generating the
teleported output state. For perfect EPR correlations, the teleported state
has unit fidelity with the original unknown input state, as can be verified
by ``Victor'' who both generates the original input and measures the
teleported output. Of course, the limit of this ideal case is unattainable
in any laboratory setting. This necessitates the introduction of operational
criteria to gauge the success of the protocol, as discussed in Refs.\
\cite{Braunstein00,Braunstein01}, and as will be applied in relation to our
experiment.

The paper is organized as follows. In Sec.\ \ref{theory} we discuss the
fidelity for quantum teleportation in the presence of losses and phase
fluctuations, including importantly for the EPR beams. In Sec.\
\ref{exp_details} this model is connected to the laboratory via a detailed
discussion of the generation of our EPR resource, including specifications of
the optical parametric oscillator (OPO) parameters, the obtainable squeezing,
and the characterization of the EPR state. The technical details of the actual
implementation of the quantum teleportation protocol are discussed in depth in
Sec.\ \ref{QT_results} with emphasis on the phase-lock servo systems and the
calibration of the classical information channels. Here, we also present new
data on the teleportation of coherent states of light. These experimental data
are compared to theoretical calculations based on the relevant parameters for
the experiment, with each parameter measured in absolute terms without
adjustment. Finally, we collect our conclusions in Sec.\ \ref{conclusion},
together with an outlook for future progress.

\section{Theory}
\label{theory}
In this section a theoretical description of the quantum
teleportation protocol for continuous variables is given. This is a
generalization of previous work in order to include all relevant detector
inefficiencies and phase offsets for the experiment. The discussion is divided
into two parts: in Sec.\ \ref{Fidelity_vis} the effect of nonideal homodyne
detectors is investigated while Sec.\ \ref{phase_qt} concerns phase
fluctuations due to imperfect phase-lock servos. Both effects turn out to be
of substantial importance in trying to accurately model experimental data.

\subsection{Detection inefficiencies}
\label{Fidelity_vis}

\begin{figure}[htb]
\begin{center}
\includegraphics[width=8.cm]{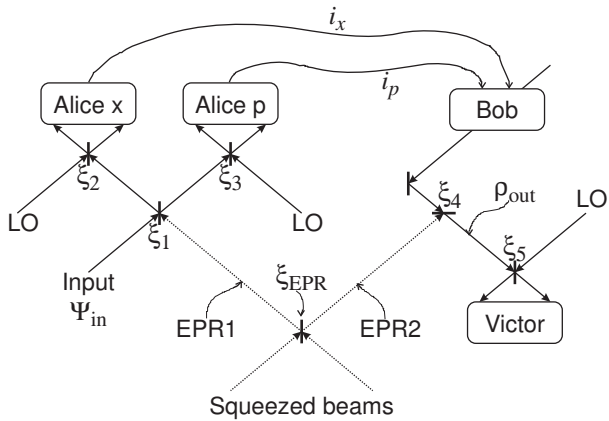}
\end{center}
\caption{Main parts in the teleportation protocol for continuous variables.
Indicated are the relevant efficiencies $(\xi_{1 \rightarrow5}, \xi_{EPR})$
that limit the teleportation fidelity.}
\label{setup_simple}
\end{figure}

Fig.\ \ref{setup_simple} shows a simplified schematic of the experiment for
teleportation of an unknown quantum state provided by the verifier Victor and
characterized by a pure input state $\left\vert \psi_{in}\right\rangle$. The
process is as follows: Alice performs measurements of the two quadratures $x$
and $p$ of the fields obtained by combining the unknown input state with EPR1.
This is done by implementing two balanced homodyne detectors where the signal
fields are each combined with a strong coherent local oscillator (LO) and the
resulting output intensities are measured. Subtracting the two photocurrents
from a given set of detectors results in a signal proportional to the
quadrature amplitude, with the relevant quadrature selected by the phase of
the LO. The efficiency of the homodyne detectors can be characterized by the
visibilities $(\xi_{2},\xi_{3})$ of the overlap between the LOs and the output
beams from Alice's beamsplitter, as well as the detectors' quantum efficiency
$(\alpha)$. Furthermore, the visibility $(\xi_{1})$ of the overlap between the
input state and EPR1 is relevant.

Because of the nature of the EPR correlations, the effect of
Alice's quadrature measurements is to project EPR2 onto a state that differs
from the unknown input state only by a phase space displacement. The necessary
displacement, however, depends on the outcome of Alice's measurements. Hence
the task for Bob is to perform this phase space displacement with the
classical information received from Alice by way of the photocurrents
$(i_{x},i_{p})$ shown in Fig.\ \ref{setup_simple}\@. In practice this is
accomplished by
overlapping EPR2 with a phase and amplitude modulated coherent state on a 99/1
beamsplitter. This modulation is directly driven (with suitable gain and phase
compensation) by the photocurrents $(i_{x},i_{p})$ from Alice's detectors. The
relevant efficiency is the visibility $(\xi_{4})$ between EPR2 and the
modulated coherent state. Finally, the quality of the teleportation can be
checked by a third party (Victor) that performs homodyne detection on the
output state. The visibility of Victor's homodyne detector is denoted $\xi
_{5}$.

As discussed in more detail in Refs.\ \cite{Braunstein00,Braunstein01}, the
performance of the teleportation protocol can be quantified by the fidelity
$F$, which is defined by
\begin{equation}
F=\left\langle \psi_{in}\left\vert \rho_{out}\right\vert \psi_{in}
\right\rangle ,
\end{equation}
which is simply the overlap between the input state $\left\vert \psi_{in}
\right\rangle$ (which is assumed to be a pure state) and the output
state characterized by a density matrix $\rho_{out}.$ In the limit of perfect
detectors (unity efficiencies) but with a finite degree of EPR correlation,
the fidelity for quantum teleportation of coherent states can be shown to be
\cite{Furusawa98,Braunstein00}
\begin{equation}
\label{fidelity_exp}
F=\frac{2}{\sigma_{Q}}\;\mbox{exp}\!\left[ -\frac{2}{\sigma_{Q}}
    \left\vert\beta_{out}-\beta_{in}\right\vert^{2}\right] ,
\end{equation}
where
\begin{subequations}
\label{Vic_variance}
\begin{align}
\sigma_{Q}  & =\sqrt{(1+\sigma_{W}^{x})(1+\sigma_{W}^{p})},\\
\sigma_{W}^{x}  & =\sigma_{W}^{p}=g^{2}+\frac{1}{2}e^{2r_{+}}(1-g)^{2}
+\frac{1}{2}e^{-2r_{-}}(1+g)^{2}.
\end{align}
\end{subequations}
Here $\sigma_{W}^{x}$ and $\sigma_{W}^{p}$ are the variances (in the Wigner
representation) of the teleported $x$ and $p$ quadratures that emerge from
Bob's beamsplitter (as shown in Fig.\ \ref{setup_simple} at $\rho_{out}$).
$g$ is gain of the classical channels, where we have assumed that the two
classical channels have the same gain and that any phase offsets have been
appropriately compensated. Furthermore, $\beta_{in}$, $\beta_{out}$ are the
amplitudes of the unknown input field and teleported output field respectively.
Finally, $r_{+},r_{-}$ are the anti-squeezing and squeezing parameters,
respectively, for the two equally squeezed beams used to produce the EPR
correlations, as will be discussed in detail in Sec.\ \ref{EPR_section}.

Any real experiment of course suffers from finite losses in propagation and
detection, with the individual efficiencies being critical due to the
fragility of quantum states of light. It turns out that the general expression
(\ref{fidelity_exp}) for the fidelity still applies to the case with losses,
but the variances of the quadratures of the teleported field generalized.
In addition, we take into account the fact that we do not observe the output
state directly, but instead measure the output photocurrent from Victor's
balanced homodyne detector. If we assume as before that the input states to
Alice are coherent states, then the quadrature variance $\sigma_{V}^{x}$
recorded by Victor for the teleported output state can be written as
\begin{align}
\label{sigma_x}
\sigma&_{V}^{x} = 1 - r_B^2\xi_4^2\xi_5^2\eta_V^2 - g_x^2\xi_1^2
                  + \frac{2g_x^2}{\xi_2^2\eta_{Ax}^2} \\
  & +\! \frac{e^{\!-2r\!_-}}{2}\!\left(g_x\xi_1
           \!\!+\! r_B\xi_4\xi_5\eta_V \right)^2
  \!+\! \frac{e^{2r\!_+}}{2}\!\left(g_x\xi_1
           \!\!-\! r_B\xi_4\xi_5\eta_V \right)^2 \nonumber ,
\end{align}
and the variance for the $p$ quadrature is given by $\sigma_{V}^{p}
=\sigma_{V}^{x}\left(g_x\rightarrow g_p, \eta_{Ax}\rightarrow\eta_{Ap},
\xi_{2}\rightarrow\xi_{3}\right)$. The fidelity is then obtained by replacing
$\sigma_W^{x,p}$ with $\sigma_V^{x,p}$ in Eqs.\ (\ref{Vic_variance}). The
non-unit reflectivity of Bob's beamsplitter appears as a loss factor $r_B$,
where in our experiment, $\left\vert r_B\right\vert^2=0.99$. $\eta_i$ are
detector efficiency factors directly related to the quantum efficiencies
$\alpha_i$ by $\alpha_i=\eta_i^2$, where the subscripts denote Alice $x$,
Alice $p$ or Victor. $g_{x,p}$ are the suitably normalized gains for the $x$
and $p$ classical channels through which Alice sends information to Bob.

In terms of the model given in Fig.\ \ref{setup_simple}, the gains $g_{x,p}$
are given explicitly by
\begin{subequations}
\label{g_explicit}
\begin{align}
g_x &= \frac{g_{x,(0)}}{\sqrt{2}} t_B \xi_2 \xi_5 \eta_{Ax} \eta_V,\\
g_p &= \frac{g_{p,(0)}}{\sqrt{2}} t_B \xi_3 \xi_5 \eta_{Ap} \eta_V.
\end{align}
\end{subequations}
Here, $g_{x,(0)}$ and $g_{p,(0)}$ are dimensionless gains that account for the
translation of the photocurrents $i_{x,p}$ into fields by Bob's amplitude and
phase modulation, where the point of reference is immediately before his
beamsplitter, which is taken to have amplitude reflection and transmission
coefficients $(r_B,t_B)$. Note that the formal limit of a phase-space
displacement by Bob is achieved only for the case $(t_B\rightarrow 0,
g_{x,(0)}\rightarrow \infty)$, with the product $t_B g_{x,(0)}$ held constant.

The convention that we adopt for the normalization of the gains $g_{x,p}$ in
Eqs.\ (\ref{sigma_x}) and (\ref{g_explicit}) is such that $g_x=g_p=1$ results
in $\beta_V=\beta_{in}$, and hence reflects an optimal reconstruction of the
input state for any sensible values of the squeezing parameters $r_{\pm}$.
The caveat here is that since we measure Victor's photocurrent and not the
field emerging from Bob, we effectively set
$\left\vert\beta_V\right\vert^2=\left\vert\beta_{in}\right\vert^2$ and not
$\left\vert\beta_{out}\right\vert^2=\left\vert\beta_{in}\right\vert^2$
as required by the protocol, where it can be easily shown that
\begin{equation}
\label{beta_V_beta_out}
\left\vert\beta_V\right\vert^2
  =\xi_5^2\eta_V^2\left\vert\beta_{out}\right\vert^2.
\end{equation}
This defect in our measurement will be discussed
quantitatively when we present our experimental data in Sec.\ \ref{infer_F}.
Note that if Victor has perfect detection efficiency ($\xi_5=\eta_V=1$),
the problem vanishes, and the result given in Eq.\ (\ref{sigma_x}) is
exact for the teleported output field emerging from Bob's beamsplitter.

The corresponding variances obtained by Alice's homodyne detectors are given
by
\begin{subequations}
\begin{align}
\label{sigma_A1}
\sigma_{A}^{x} & = 1+\frac{1}{4}\left( e^{-2r_{-}} + e^{2r_{+}} - 2\right)
                   \xi_{1}^{2}\xi_{2}^{2}\eta_{Ax}^{2},\\
\label{sigma_A2}
\sigma_{A}^{p} & = 1+\frac{1}{4}\left( e^{-2r_{-}} + e^{2r_{+}} - 2\right)
                   \xi_{1}^{2}\xi_{3}^{2}\eta_{Ap}^{2}.
\end{align}
\end{subequations}

Several limiting cases associated with these expressions are worth noting.
In the classical case where there is no EPR entanglement ($r_{+}=r_{-}=0$),
and with perfect homodyne detectors $(\xi_{1\rightarrow5}=\eta_i=1)$, we
obtain $\sigma_{V}^{x}=\sigma_{V}^{p}=3$, corresponding to three units of
vacuum noise in Victor's homodyne detector. One unit stems from the vacuum
noise intrinsic to the input coherent state, while the two extra units can
be traced back as the quantum duties added in each crossing of the border
between quantum and classical domains corresponding to Alice's quadrature
measurements and Bob's phase space displacement \cite{B&K98}. This means that
for classical teleportation of coherent states, the best achievement possible
is reconstructing the input state with two extra units of vacuum noise added
\cite{Braunstein00,Braunstein01}. The three vacuum units correspond to
excess noise recorded in Victor's homodyne detector of $4.77\:\mbox{dB}$
above the vacuum-state limit for his detector. With quantum entanglement
it is possible to beat this limit and observe noise reduction below the
$4.77\:\mbox{dB}$ level in Victor's detector. The measured noise reduction
can then be transferred into a fidelity through Eq.\ (\ref{fidelity_exp}).
As analyzed in Refs.\ \cite{Braunstein00,Braunstein01}, the classical
boundary for teleportation of coherent states is $F=0.5$.

In the case of nonideal detectors, $g_x=g_p=1$ still preserves optimal
teleportation for the normalized gains, in the sense that $\beta_V=\beta_{in}$.
However, the normalization is performed by effectively tuning the
unnormalized gains $g_{(0)}$ by
\begin{equation}
\label{g_nonideal}
g_{x,(0)}^{\mathit{nonideal}} \longrightarrow \left(\xi_2\xi_5\eta_{Ax}\eta_V
  \right)^{-1} g_{x,(0)}^{\mathit{ideal}},
\end{equation}
and similarly for $p$. Thus in the nonideal case, the actual gain is larger
than in the ideal case, reflecting the fact that the gain must now compensate
for Alice's and Victor's detection losses in order to ensure $\beta_{V} =
\beta_{in}$. As a consequence, the fidelity drops below $F=0.50$ with no
entanglement ($r_{\pm}=0$). In our experiment, the detection efficiencies
are characterized by the measured visibilities and quantum efficiencies,
which in the best case are given by $\xi_{1}=0.986$, $\xi_{2}=\xi_{3}=0.995$,
$\xi_{4}=0.988$, $\xi_{5}=0.985$, and $\alpha_V=\alpha_{Ax}=\alpha_{Ap}=
0.988$. With these experimentally achievable efficiency factors, we find that
$\sigma_V^{x,p}=4.84\:\mbox{dB}$ and $F=0.494$ when $r_{\pm}=0$.

\begin{figure}[htb]
\begin{center}
\includegraphics[width=8.cm]{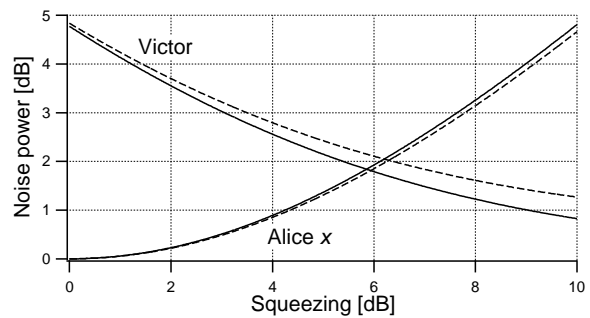}
\caption{Noise powers in dB above the vacuum-state limit for Alice's $x$
detector and for Victor's detector as a function of the degree of squeezing
of each squeezed vacuum state constituting the EPR state. The solid traces
are for an ideal case where both Alice and Victor have perfect detection
efficiency and all relevant beams are perfectly overlapped, that is,
$\xi_{1\rightarrow 5}=\alpha_i=1$. The dashed traces show the noise levels
for a real (nonideal) case where the visibilities
correspond to the experiment described below and are $\xi_{1}=0.986$,
$\xi_{2}=\xi_{3}=0.995$, $\xi_{4}=0.988$, $\xi_{5}=0.985$, and the quantum
efficiencies of photodetectors are $\alpha_i=0.988$. The squeezing
given in the figure is the squeezing just before the EPR beamsplitter.}
\label{Vic_A1}
\end{center}
\end{figure}

Fig.\ \ref{Vic_A1} shows the excess noise recorded by Victor and Alice $x$
(or equivalently Alice $p$) as a function of the amount of squeezing, both
for the ideal case with perfect detection efficiencies, and for the nonideal
case with detector efficiencies given above. With no squeezing in the ideal
case, we see from the solid curves that Victor obtains exactly
$4.77\:\mbox{dB}$ of excess noise as expected and as discussed above while
Alice is shot-noise-limited. With imperfect detection efficiencies as shown
by the dashed curves, Alice remains shot-noise-limited, while Victor records
excess noise higher than $4.77\:\mbox{dB}$. In fact, the only relevant
efficiencies that drive Victor's recorded noise above $4.77\:\mbox{dB}$
involve Alice's homodyne detectors, namely $(\xi_2,\eta_{Ax})$ for the $x$
quadrature, and $(\xi_3,\eta_{Ap})$ for the $p$ quadrature. All other
detection losses can be compensated by the gains $g_{x,p}$ when $r_{\pm}=0$.

As the squeezing is increased so that now $r_{\pm}>0$, Victor records noise
reduction below the $r_{\pm}=0$ level. By contrast, Alice's noise increases
above the vacuum level at her detectors, and in the limit of infinite
squeezing, Alice's noise diverges while Victor's excess noise is suppressed
to the vacuum level. Notice that with perfect detection efficiencies,
$\sigma_V^{x,p}<4.77\:\mbox{dB}$ for any $r_{\pm}>0$. With imperfect
efficiencies, this is not true. In effect, some of the squeezing is ``wasted''
to compensate for the nonideal efficiencies. Since our experimental
visibilities are close to unity, this loss can be neglected as it is below
the level of other experimental uncertainties for small values of $r_\pm$.
However, with large degrees of squeezing, the disparity between the ideal
and nonideal cases increases and cannot be ignored, as can be seen from Fig.\
\ref{Vic_A1}. The reason for this trend is that now the visibilities $\xi_1$
and $\xi_4$ that characterize the overlap of the EPR beams with Alice's and
Bob's relevant beams, as well as the non-unit reflectivity $r_B$ of Bob's
beamsplitter, become important. The losses from non-unit $\xi_1$ and $\xi_4$
obviously cannot be compensated by the gains of the classical channels.

\begin{figure}[htb]
\begin{center}
\includegraphics[width=8.cm]{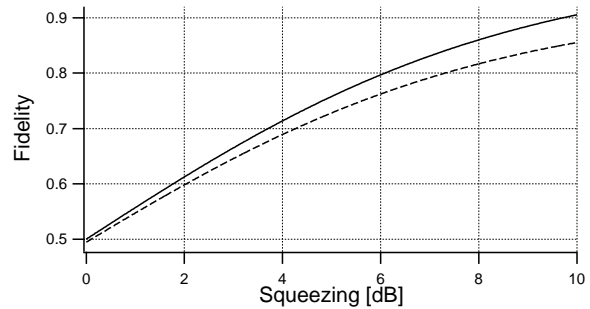}
\caption{The fidelity for Victor's teleported output as a function of the
degree of squeezing with same parameters as in Fig.\ \ref{Vic_A1}. Again,
the solid trace describes a perfect case with ideal detectors, while the
dashed trace describes the imperfect case as relevant to our experiment.}
\label{Fidelity_squeezing}
\end{center}
\end{figure}

The noise reduction at Victor can be transferred into a teleportation
fidelity, with the result plotted in Fig.\ \ref{Fidelity_squeezing}. The solid
and dashed curves for the ideal and nonideal cases mimic the conclusions
discussed above for the variances $\sigma_V^{x,p}$ measured by Victor.

\begin{figure}[htb]
\begin{center}
\includegraphics[width=8.cm]{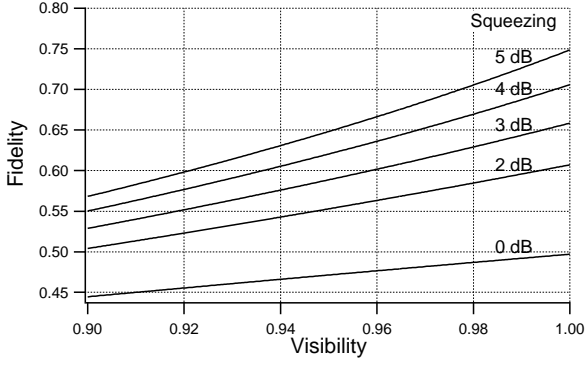}
\caption{Fidelity as a function of visibility for different values of
the degree of squeezing. We have assumed $\xi_{1\rightarrow5}=\xi$ and
$\alpha=0.988$.}
\label{Fidelity_visibility}
\end{center}
\end{figure}

Of course, the teleportation fidelity is very dependent on the detector
efficiencies. We investigate this point in more detail in Fig.\
\ref{Fidelity_visibility}, where the fidelity is plotted as a function of a
single global visibility $\xi$\ (assuming $\xi_{1\rightarrow5}=\xi$) and
where the quantum efficiencies of all the photodetectors are $\alpha=0.988$.
This figure clearly illustrates the need for a high amount of squeezing as
well as very efficient spatial mode-matching of our optical beams
to achieve high fidelity quantum teleportation.

\subsection{Phase fluctuations}
\label{phase_qt}
Not only losses associated with the detection efficiencies limit the achieved
fidelity for quantum teleportation. Also the quality of the servo-control
systems that lock various phases (e.g.,\ the local oscillator phases at
Alice's detectors) appear to be of significant importance, since phase
deviations due to nonideal locking turn out to deteriorate the noise reduction
measured by Victor. We will see that this mathematically corresponds to mixing
in terms proportional to the anti-squeezed quadratures of the squeezed beams
constituting the EPR state.

In a realistic model of the experiment we include phase offsets of four servo
locks: the EPR lock, Alice's two homodyne detectors, as well as Bob's lock of
the phase between the second EPR field and the classical field. The analysis
presented here will be a straightforward generalization of the derivation in
\cite{vanLoock00} based on the Heisenberg picture. The quadratures of the two
EPR fields (1 and 2) are obtained by combining two squeezed fields with the
angle between the squeezing ellipses equal to $\pi/2.$ Although we have
investigated a more complete model, here we account for the phase deviation
away from $\pi/2$ by introducing an angle offset $\theta_{E}$ for field 2. In
this simple nonideal case, we obtain for the fields emerging from the EPR
beamsplitter
\begin{subequations}
\label{x12_p12}
\begin{align}
\hat{x}_{1,2} & =\frac{1}{\sqrt{2}}\left( e^{r_{+}}\hat{x}_{1}^{(0)}\mp
\cos\theta_{E}e^{-r_{-}}\hat{x}_{2}^{(0)}\mp\sin\theta_{E}e^{r_{+}}\hat{p}
_{2}^{(0)}\right) ,\nonumber\\
& \\
\hat{p}_{1,2} & =\frac{1}{\sqrt{2}}\left( e^{-r_{-}}\hat{p}_{1}^{(0)}\mp
\cos\theta_{E}e^{r_{+}}\hat{p}_{2}^{(0)}\pm\sin\theta_{E}e^{-r_{-}}\hat{x}
_{2}^{(0)}\right) ,\nonumber\\
&
\end{align}
\end{subequations}
where $\hat{x}_{1,2}^{(0)}$ and $\hat{p}_{1,2}^{(0)}$ are vacuum state
operators for $(x,p)$ for the input beams 1 and 2 to the beamsplitter,
respectively. Further downstream, EPR beam 1 is mixed with the unknown
input state on a 50/50 beamsplitter creating the modes
$\hat{u}=(\hat{a}_{in}-\hat{a}_{1})/\sqrt{2}$ and
$\hat{v}=(\hat{a}_{in}+\hat{a}_{1})/\sqrt{2},$ and Alice measures the two
quadrature amplitudes of the corresponding state in her homodyne detectors.
Allowing for small phase deviations $\theta_{Ax}$ and $\theta_{Ap}$ in the
detection process, we find that the quadratures measured by Alice now become
\begin{subequations}
\label{xu_pv}
\begin{align}
\hat{x}_{u}(\theta_{Ax})  & =\hat{x}_{u}\cos\theta_{Ax}+\hat{p}_{u}\sin
  \theta_{Ax},\\
\hat{p}_{v}(\theta_{Ap})  & =\hat{p}_{v}\cos\theta_{Ap}-\hat{x}_{v}\sin
  \theta_{Ap}.
\end{align}
\end{subequations}
Finally, Bob performs a phase space displacement of the second EPR beam by
overlapping with the coherent beam containing the classical information
received from Alice. Allowing again for a phase offset $\theta_{B}$ in Bob's
phase-space displacement, we calculate that the quadrature operators for the
teleported state exiting the apparatus for investigation by Victor is given by
\begin{subequations}
\begin{align}
\hat{x}_{V}  & =\hat{x}_{2}\cos\theta_{B}+\hat{p}_{2}\sin\theta_{B}+\sqrt{2}
\hat{x}_{u}(\theta_{Ax}),\\
\hat{p}_{V}  & =\hat{p}_{2}\cos\theta_{B}-\hat{x}_{2}\sin\theta_{B}+\sqrt{2}
\hat{p}_{v}(\theta_{Ap}),
\end{align}
\end{subequations}
where the normalized gains of the classical channels have been taken to be
unity. Using Eqs.\ (\ref{x12_p12}) and (\ref{xu_pv}), we arrive at
expressions for the Heisenberg operators for the teleported field received
by Victor, namely
\begin{widetext}
\begin{subequations}
\begin{align}
\sqrt{2}\hat{x}_{V} =
 &   \left( \cos\theta_B - \cos\theta_{Ax}\right) e^{r_+}\hat{x}_1^{(0)}
   + \left( \sin\theta_B - \sin\theta_{Ax}\right) e^{-r_-}\hat{p}_1^{(0)}
       \nonumber\\
 & + \left[ \cos\theta_E(\cos\theta_B + \cos\theta_{Ax})
          - \sin\theta_E(\sin\theta_B + \sin\theta_{Ax})\right]
       e^{-r_-}\hat{x}_2^{(0)} \nonumber\\
 & + \left[ \sin\theta_E(\cos\theta_B + \cos\theta_{Ax})
          + \cos\theta_E(\sin\theta_B + \sin\theta_{Ax})\right]
       e^{r_+}\hat{p}_2^{(0)}
     + \sqrt{2}\cos\theta_{Ax}\hat{x}_{in}
     + \sqrt{2}\sin\theta_{Ax}\hat{p}_{in}, \\
 & \nonumber \\[-6pt]
\sqrt{2} \hat{p}_{V} =
 & - \left( \sin\theta_B + \sin\theta_{Ap}\right) e^{r_+}\hat{x}_1^{(0)}
   + \left( \cos\theta_B + \cos\theta_{Ap}\right) e^{-r_-}\hat{p}_1^{(0)}
        \nonumber\\
 & + \left[ \sin\theta_E(\cos\theta_{Ap} - \cos\theta_B)
          + \cos\theta_E(\sin\theta_{Ap} - \sin\theta_B)\right]
       e^{-r_-}\hat{x}_2^{(0)} \nonumber\\
 & + \left[ \sin\theta_E(\sin\theta_{Ap} - \sin\theta_B)
          + \cos\theta_E(\cos\theta_{B} - \cos\theta_{Ap})\right]
       e^{r_+}\hat{p}_2^{(0)}
     + \sqrt{2}\cos\theta_{Ap}\hat{p}_{in}
     - \sqrt{2}\sin \theta_{Ap} \hat{x}_{in}.
\end{align}
\end{subequations}
\end{widetext}

By utilizing these expressions, the variances of the two quadratures measured
by Victor can be calculated. Assuming the phase excursions are small, we
expand to lowest order. We recall that the aim of this calculation is to
describe the impact of phase fluctuations in the various servo-controls. Hence
we assume that there are no static offsets (which we believe our current
procedures adequately null), so that all the phase excursions vanish on
average, $\overline{\theta}=0,$ and only deviations expressed by the second
order moments contribute. Furthermore, it is assumed that all the phase
fluctuations are independent, so that products of phases vanish on average.

After some algebra, we finally arrive at
\begin{subequations}
\label{vic_phases}
\begin{align}
\sigma_V(x) = \left<\Delta \hat{x}_{V}^2 \right> =&
  1 + \left[2 - \frac{1}{2}\overline{\theta_{Ax}^2}
    - \frac{1}{2}\overline{\theta_B^2}
    - 2 \overline{\theta_E^2} \right] e^{-2 r_{-}} \nonumber\\
& + \left[\frac{1}{2}\overline{\theta_{Ax}^2}
    + \frac{1}{2}\overline{\theta_{B}^2}
    + 2 \overline{\theta_E^2} \right] e^{2 r_{+}},\\
\sigma_V(p) = \left<\Delta \hat{p}_{V}^2 \right> =&
  1 + \left[2 - \frac{1}{2}\overline{\theta_{Ap}^2}
    - \frac{1}{2}\overline{\theta_B^2} \right] e^{-2 r_{-}} \nonumber\\
& + \left[\frac{1}{2}\overline{\theta_{Ap}^2}
    + \frac{1}{2}\overline{\theta_B^2} \right] e^{2 r_{+}},
\end{align}
\end{subequations}
where the various $\overline{\theta_{i}^{2}}$ are meant to be
associated with the residual RMS fluctuations arising from the nonideal
performance of our locking servos. Explicit dependence on the phase
$\theta_V$ of Victor's LO is given by
\begin{equation}
\sigma_V[x(\theta_V)] = \sigma_V(x) \cos^2{\theta_V}
                      + \sigma_V(p) \sin^2{\theta_V}.
\end{equation}

These equations make quantitative the obvious intuition that the effect of the
phase fluctuations is to add extra noise in the quadratures measured by Victor
through components proportional to the anti-squeezed quadrature. In fact,
relatively small phase fluctuations ($\sim 1^{\circ}$ RMS) can
degrade the noise reduction that would otherwise have been recorded by Victor,
and consequently also the achieved fidelity. 

From these equations, we see that particular phase fluctuations 
contribute in quite
different ways. Phase fluctuations at Bob contribute equally to excess noise
in the $x$ and $p$ quadratures and will consequently be seen as a constant
shift in the noise measured in Victor's homodyne detector while scanning the
local oscillator. The same effect is found from fluctuations in the locking
of the local oscillator phases at Alice $x$ and Alice $p$ provided
$\overline{\theta_{Ax}^{2}}=\overline{\theta_{Ap}^{2}}.$ However, phase
fluctuations in the EPR lock are seen to modify Victor's $x$ and $p$
quadratures differently and therefore imply modulation of the noise measured
by Victor. The relevant second order moments $\overline{\theta_{i}^{2}}$ for
the various locks can be obtained experimentally by measuring the RMS noise
of the error signals in locked operation. Typically measurements give
$\sqrt{\overline{\theta_{i}^{2}}} \simeq 2$ to $6\:\mbox{degrees}.$ From
Eqs.\ (\ref{vic_phases}) it is seen that fluctuations in the phase with which
the squeezed beams are combined to form the EPR beams are most critical since
$\overline{\theta_{E}^{2}}$ contributes with a coefficient four times higher
than the other phase terms to the mixing with the anti-squeezing term.

\begin{figure}[htb]
\begin{center}
\includegraphics[width=8.cm]{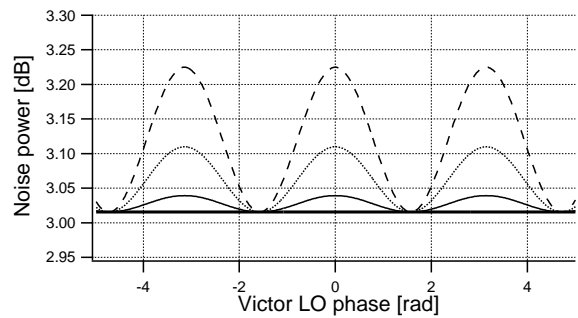}
\caption{Noise power recorded by Victor's balanced detector when scanning his
local oscillator for $\sqrt{\overline{\theta_{E}^{2}}}
=0,2,4,6\:\mbox{degrees}$ (corresponding to the bold, thin, dotted, and dashed
curves, respectively), and for $\overline{\theta_{Ax}^{2}}=\overline
{\theta_{Ap}^{2}}=\overline{\theta_{B}^{2}}=0$. Realistic values of the
degrees of squeezing ($-3\:\mbox{dB}$) and of anti-squeezing ($7\:\mbox{dB}$)
have been used.}
\label{QT_phases}
\end{center}
\end{figure}

Fig.\ \ref{QT_phases} shows the calculated noise in Victor's
homodyne detector for different levels of phase fluctuations in the EPR lock
employing realistic values of squeezing and anti-squeezing for the experiment
discussed in the following sections. The modulation of Victor's signal is seen
to be up to about $0.2\:\mbox{dB}$ peak to peak which turns out to imply a
significant reduction of the achieved fidelity.

\section{Generation of the quantum resource}
\label{exp_details}
\begin{figure}[h]
\begin{center}
\includegraphics[width=8.cm]{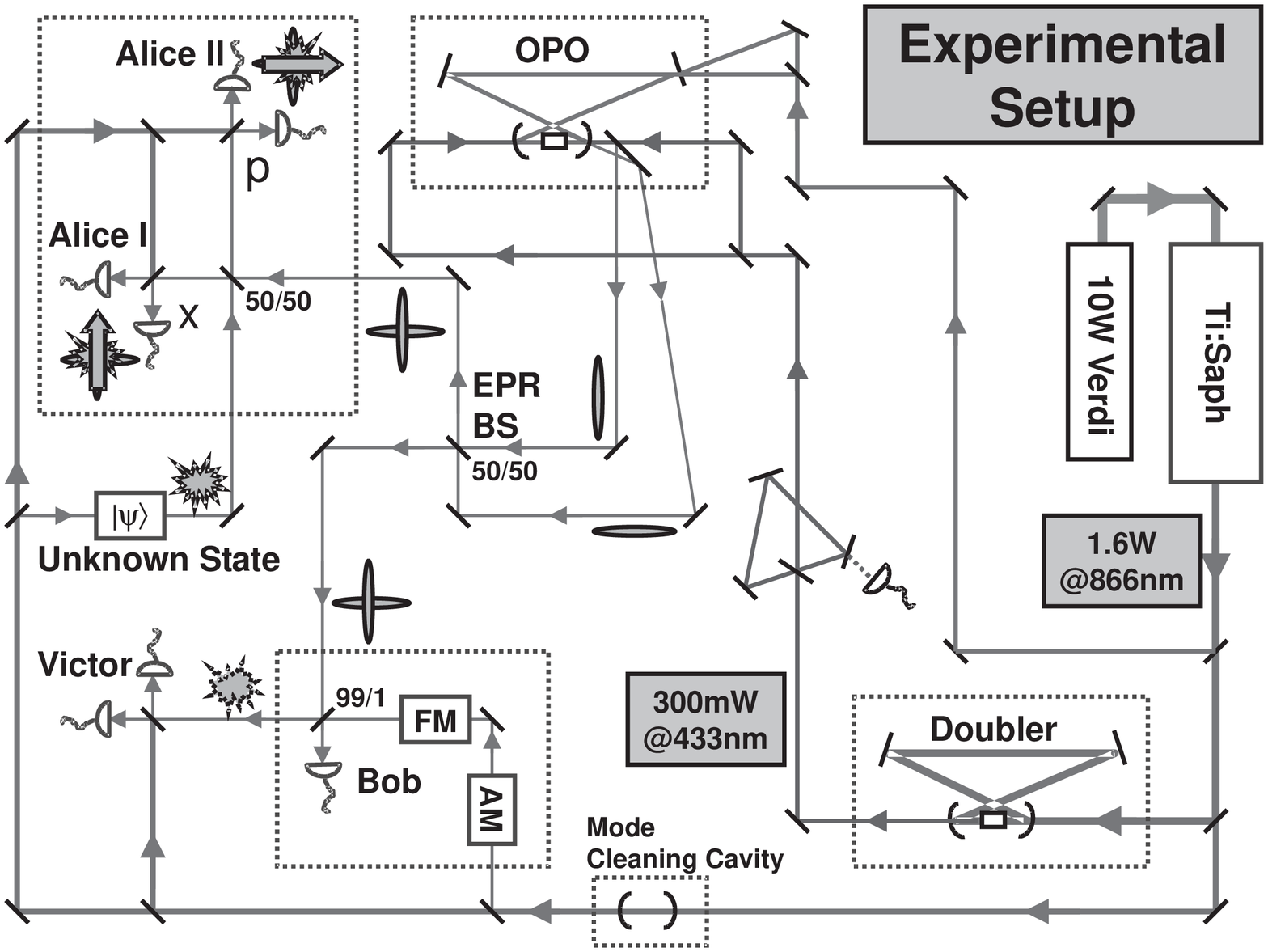}
\caption{Sketch illustrating the basic optical parts in the quantum
teleportation experiment. See the main text for a careful discussion.}
\label{setup_detailed}
\end{center}
\end{figure}

A more complete figure of the experimental setup is given in Fig.\
\ref{setup_detailed}. A $10\:\mbox{W}$ Verdi was used to pump a single
frequency Ti:Sapphire laser operating at $866\:\mbox{nm}.$ This laser system
provided about $1.6\:\mbox{W}$ of IR. About $80\%$ to $85\%$ of this light was
sent to an external frequency doubler to generate an efficient $433\:\mbox{nm}$
pump source for the optical parametric oscillator (OPO). Typically about
$300\:\mbox{mW}$ of blue light was produced that could be modematched
to the OPO using a triangular ring cavity. Furthermore, the pump was divided
into two beams, which allowed pumping the OPO from two directions to produce
two independent squeezed beams. A detailed description of this setup for
generation of highly squeezed light can be found in Ref.\ \cite{Polzik92}.
About $10\%$ of the IR light from the Ti:Sapphire laser was spatially
filtered in a mode cleaning cavity and used down stream in the experiment for
locking the OPO, as local oscillators in the homodyne detectors, for Bob's
displacement beam, and as the input coherent state for the actual
teleportation. Combining the two squeezed beams on a 50/50 beamsplitter with
the phases locked so that the squeezing ellipses are perpendicular to each
other, the EPR state was generated which is the quantum resource necessary for
the actual quantum teleportation protocol described previously in relation to
Fig.\ \ref{setup_simple}. In the current section a detailed description of the
generation of the EPR state is given with a careful characterization of both
classical and quantum properties of the OPO. The actual implementation of the
full teleportation protocol follows in Sec.\ \ref{QT_results}.

\subsection{Loss and gain in the OPO}

The OPO cavity was a bow-tie ring configuration consisting of two curved
mirrors (radius of curvature $5\:\mbox{cm})$ and two plane mirrors. The total
cavity length was $48\:\mbox{cm}$. The focus positioned between the two curved
mirrors had a waist size of $21\:\mu\mbox{m}$ where the $1\:\mbox{cm}$ long
nonlinear potassium niobate ($\mbox{KNbO}_{3}$) crystal was positioned. The
use of a-cut potassium niobate allowed noncritical temperature phase-matching
of a degenerate parametric process. To generate a pump for the OPO, the output
from the Ti:Sapphire laser was frequency doubled in another external cavity
also using potassium niobate as the nonlinear medium \cite{Polzik91}. In this
way $300\:\mbox{mW}$ of pump light at $433\:\mbox{nm}$ was generated.

In the OPO nonlinear down-conversion transformed energy from the pump field at
$433\:\mbox{nm}$ into the parametric field at $866\:\mbox{nm}.$ In the current
application the OPO was only driven below oscillation threshold, where
spontaneous parametric emission can produce a squeezed vacuum state. The
parametric field was resonant in the OPO while the pump light from the
frequency doubler was divided into two beams, each of which was used in
single pass of the OPO crystal from a counter propagating direction.

The OPO performance can be characterized once specifying the output coupler
intensity transmission $T,$ the effective nonlinearity $E_{NL},$ and the
intracavity round-trip loss $\mathcal{L}.$ In the current experiment the
output coupler transmission was fixed at $T=10\%$ which was chosen to optimize
squeezing. The total information about the nonlinear interaction can be
captured in the single parameter ${E}_{NL}$ that depends on the focusing,
length of the crystal, phase-matching and crystal properties. It can
operationally be defined as ${E}_{NL}=P_{2}/P_{1}^{2},$ where $P_{2}$ is the
second harmonic power generated in single pass frequency doubling of a
fundamental pump $P_{1}.$ In the current setup we measured
$E_{NL}=0.021W^{-1}.$ 

\begin{figure}[h]
\begin{center}
\includegraphics[width=8.cm]{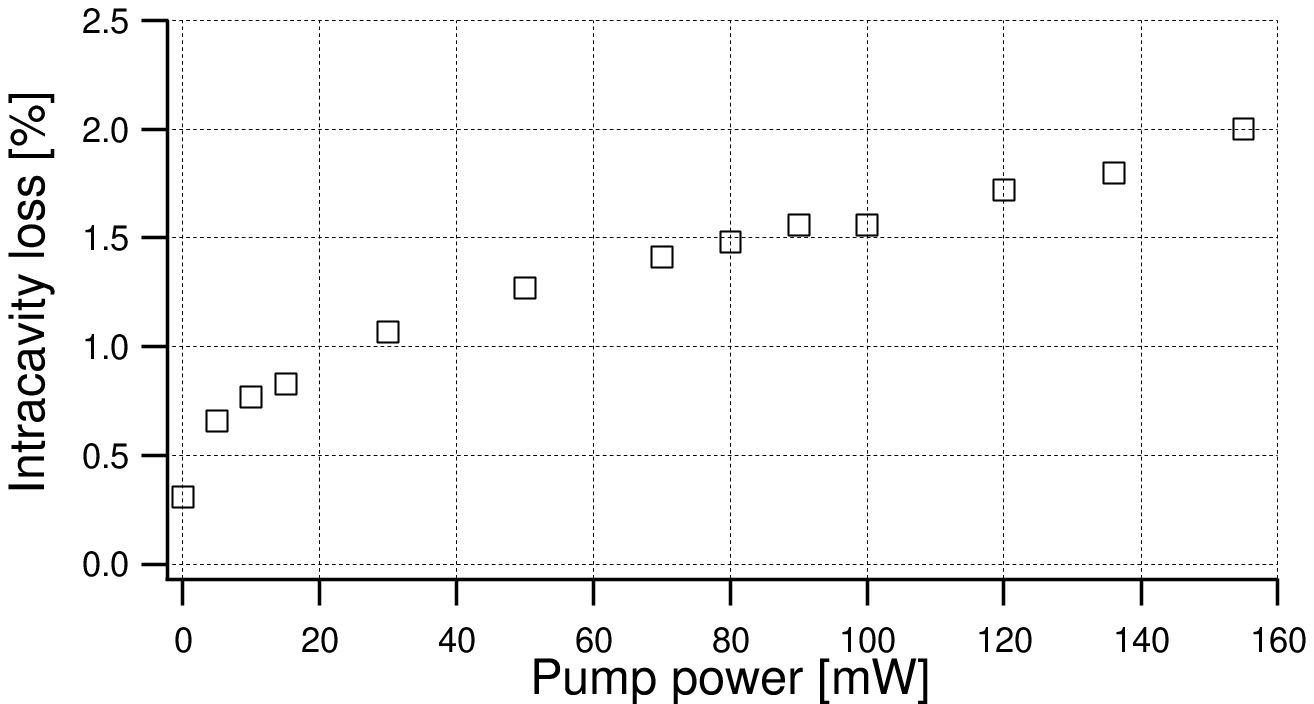}
\end{center}
\caption{Measured total intracavity round-trip cavity loss $\mathcal{L}$
as a function of the blue pump $P_{2}.$ Note that the transmission $T$
of the output coupler is not included.}
\label{loss}
\end{figure}

Contributing to the intracavity loss are nonideal
antireflection coatings of the potassium niobate as well as leakage from the
three high-reflection coated cavity mirrors. This passive loss was measured to
be $\mathcal{L}_{p}=0.3\%.$ Unfortunately potassium niobate also suffers from
an inherent loss mechanism that adds to the passive losses \cite{Mabuchi94,
Shiv95}. This nonlinear loss arises in the OPO in the presence of the blue
pump beam and has been termed blue-light-induced infrared absorption (BLIIRA).
It is believed to originate from impurities in the crystal and is found to
vary substantially from crystal sample to sample. At a high pump level of the
OPO, BLIIRA turns out to be the dominating loss mechanism and eventually
becomes the limiting factor for the amount of squeezing obtained. The losses
$\mathcal{L}_{b}$ due to BLIIRA could be monitored in the OPO by measuring the
reflection dip of the injected pump beam while scanning the cavity around
resonance. Typical measurements of the total intracavity loss
($\mathcal{L} = \mathcal{L}_{p} + \mathcal{L}_{b}$) are shown in
Fig.\ \ref{loss} as a function of the blue pump power. In this case the
OPO was only pumped along one direction. We observe that the total
loss increases up to about $2\%$ at the highest pump level of $P_2 =
155\:\mbox{mW}$.

\begin{figure}[h]
\begin{center}
\includegraphics[width=8.cm]{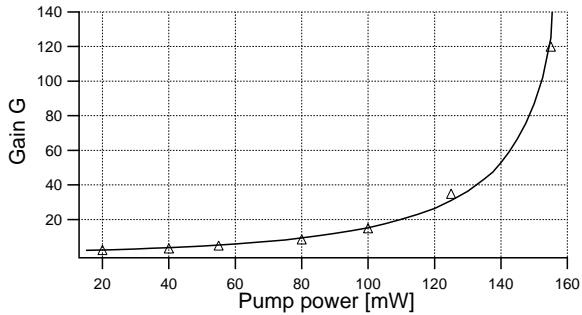}
\end{center}
\caption{Measured gain (points) and corresponding theoretical curve based on
the measured loss and nonlinearity.}
\label{gain}
\end{figure}

The OPO could also be operated as a phase sensitive amplifier as a way to test
the classical performance of the device. For that purpose an $866\:\mbox{nm}$
beam was seeded into the OPO and by scanning the injection phase slowly, the
amplification factor $G$ was measured. When injecting the seed beam through
the small transmission of a high reflection mirror and measuring the
amplification of the light through the output coupler mirror, the gain is
given by
\begin{equation}
G=\frac{1}{\left(  1-\sqrt{P_{2}/P_{2,t}}\right)  ^{2}},
\end{equation}
where $P_{2,t}$ is the oscillation threshold of the OPO given by
\begin{equation}
P_{2,t}=\frac{(T+\mathcal{L})^{2}}{4E_{NL}}.
\end{equation}
Fig.\ \ref{gain} shows measured values of the gain as a function of pump power
for single-sided pump of the OPO as well as the theoretical curve based on the
measured loss and nonlinearity. The gain diverges as approaching threshold and
from these data we estimate $P_{t}\simeq 190\:\mbox{mW}$. The agreement
between experiment and theory is apparently quite good, here with no
adjustable parameters.

\subsection{Squeezing}

The phase-sensitive amplification in the OPO can be exploited for generating
squeezed states of light. In this case the two input vacuum noise quadratures
are amplified and deamplified, respectively, creating a squeezed vacuum state.
Balanced homodyne detection allows phase sensitive detection of the quantum
noise of the squeezed state. With this method the signal field is overlapped
with a strong coherent local oscillator (LO) on a 50/50 beamsplitter. The two
output beams from the beamsplitter are measured and the corresponding
photocurrents subtracted. In this way the weak quantum noise of the signal is
amplified to achieve a signal substantially above the thermal noise floor of
the photodiodes. The photodiodes used were a special part made by Hamamatsu
with a measured quantum efficiency $\alpha = 98.8 \pm1.0 \%.$

\begin{figure}[htb]
\begin{center}
\includegraphics[width=8.cm]{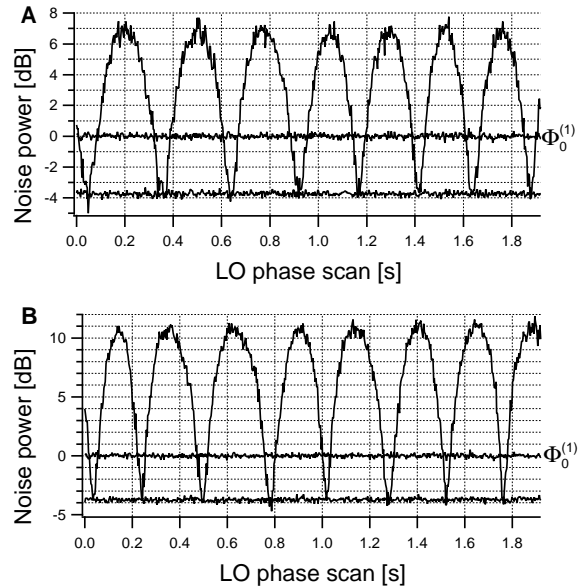}
\end{center}
\caption{Typical squeezing traces recorded by homodyne detection while scanning
the phase of the LO. The measurement frequency was $1.475\: \mbox{MHz}$ within
a resolution bandwidth of $30\: \mbox{kHz}$ and with a video bandwidth of
$300\:\mbox{Hz}.$ The flat traces at $0\:\mbox{dB}$ are the respective vacuum
levels $\Phi_{0}^{(1)}$, taken with a $5$ trace average. Also displayed are
the flat traces that correspond to the minimum noise level attained, again
with a 5 trace average. These traces were obtained by locking the LO
phase to the squeezed quadrature and lead to estimates of the squeezing and
anti-squeezing of: (A) $-3.73\:\mbox{dB}$ and $6.9\:\mbox{dB}$ with $42\:
\mbox{mW}$ pump power; and (B) $-3.73\:\mbox{dB}$ and $10.8\:\mbox{dB}$ with
$107\:\mbox{mW}$ pump power. In both cases, the OPO was pumped from a single
direction. Victor's detection efficiency was characterized by a homodyne
visibility $\xi=0.972$ and photodiode quantum efficiency $\alpha=0.988$.}
\label{squeez_trace}
\end{figure}

Two typical squeezing traces at different pump levels are presented in
Fig.\ \ref{squeez_trace}. They were obtained by recording Victor's noise
power with the spectrum analyzer when scanning
the phase of the LO. We observe the phase sensitive noise with the maximum
and minimum corresponding to measuring the anti-squeezed and squeezed
quadratures, respectively. With the signal beam blocked, the vacuum state
level $\Phi_{0}^{(1)}$ was recorded, and squeezing corresponds to noise
reduction below this level. By locking the LO phase to the squeezed
quadrature, we obtained the flat traces below $\Phi_{0}^{(1)}$, as
shown in Fig.\ \ref{squeez_trace}.
We infer squeezing of $3.73\: \mbox{dB}$ below the vacuum level and
anti-squeezing of $6.9\: \mbox{dB}$ with pump power of $42\:\mbox{mW}$.
At the higher pump level shown ($107\:\mbox{mW}$), the degree of squeezing
remained at $3.73\: \mbox{dB}$ below the vacuum level while the degree of
anti-squeezing increased to $10.8\:\mbox{dB}$.

In the actual teleportation experiment the fidelity is ultimately limited by
the amount of squeezing available. However, as discussed in Section
\ref{phase_qt}, in a non-perfect experiment the amount of anti-squeezing
is also important. In that section, we concluded that fluctuations
in the servo locks will degrade the fidelity with contributions from the
anti-squeezed quadratures. For that reason it is important to find the optimum
operation point of the OPO where the degree of squeezing is large, while at
the same point, the anti-squeezing has not grown too large. Such a
compromise is made necessary by the BLIIRA, which limits the degree of quantum
noise reduction in a power-dependent fashion, while the noise from the
anti-squeezed quadrature continues to grow.

\begin{figure}[h]
\begin{center}
\includegraphics[width=8.cm]{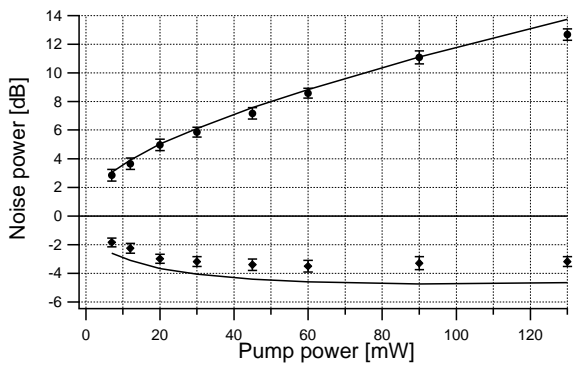}
\end{center}
\caption{Measured squeezing (diamonds) and anti-squeezing (dots) at a
frequency offset of $1.475\:\mbox{MHz}$ as a function of the OPO pump power
when pumping from only one direction. The curves are the corresponding
theoretical results for nonlinearity $E_{NL} = 0.019 W^{-1}$ and measured
cavity loss similar to that given in Fig.\ \ref{loss}. The measured quantum
efficiency was $\alpha = 0.988$, the homodyne visibility $\xi = 0.990$, and
the propagation loss of the squeezed beam was $5.7\%.$}
\label{squeez_pump}
\end{figure}

Fig.\ \ref{squeez_pump} shows the variation of the measured
squeezing and anti-squeezing with the OPO pump power as well as the
corresponding theoretical curves based on the measured experimental parameters
(loss and nonlinearity) discussed in the previous section. The data have
been corrected for the thermal noise level of the detectors, which was
$17\:\mbox{dB}$ below the vacuum noise level for an LO of $2\:\mbox{mW}$.
The squeezing is seen to level off at about $-3.5\:\mbox{dB}$ already at a
pump of $45\:\mbox{mW}$ while the anti-squeezing increases with the pump. This
indicates that in the teleportation experiment it would be most favorable to
operate at this relatively weak pump level of the OPO and that decreased
teleportation fidelity might be expected when increasing the pump further
(e.g.,\ due to mixing in of noise from the anti-squeezed quadrature from
the imperfections in servo control discussed in Sec.\ \ref{phase_qt}).

We note that the measured squeezing is lower than predicted
from the OPO parameters discussed above; indeed we predict about
$4.7\:\mbox{dB}$ squeezing at high pump level. This discrepancy might be due to
offset fluctuations in the OPO lock as well as phase fluctuations between the
local oscillator and the squeezed beam in the homodyne detector. In favor of
such an explanation is the fact that the theory predicts the anti-squeezed
quadrature better than the squeezing quadrature, and the broad maximum from
the anti-squeezing (see Fig.\ \ref{squeez_trace}) is expected to be much less
sensitive to phase fluctuations than the narrow squeezing minimum.

The above data were taken when the OPO was only pumped from a single
direction. In order to obtain two squeezed beams necessary to generate the EPR
correlations, the OPO was pumped in two counter-propagating directions. In this
case we expect lower squeezing than from a single pumped OPO due to increased
BLIIRA. This reduction in squeezing was measured to range from less than
$0.3\:\mbox{dB}$ at total pump powers below around $80\:\mbox{mW}$, to
$0.5\:\mbox{dB}$ at higher pump powers. To estimate the degree of squeezing
in the double-pumped case,
we take the degree of squeezing obtained in Fig.\ \ref{squeez_trace} and
correct for Victor's homodyne visibility and finite
detector thermal noise, to find that in the single-pumped case, we have
$-4.1\:\mbox{dB}$ of available squeezing. Thus we estimate that we have
about $-3.6$ to $-3.8\:\mbox{dB}$ squeezing in the double-pumped case.

\subsection{EPR correlations}
\label{EPR_section}
The EPR correlated beams were generated by combining two independently
squeezed beams with the relative phase servo-locked to be $\pi/2.$ These
continuous variable EPR correlations are of the type originally discussed by
Einstein, Podolsky, and Rosen \cite{EPR}. The two output beams 1 and 2 from
the EPR beamsplitter possess correlations as expressed by the variances
$\sigma(x_{1}\pm x_{2})=2\sigma_{\pm}$ and
$\sigma(p_{1}\pm p_{2})=2\sigma_{\mp},$
where $\sigma_{+}$ and $\sigma_{-}$ are the variances of the
anti-squeezed and squeezed quadratures of the two input beams, i.e.
$\sigma_{+}>1,$ $\sigma_{-}<1,$ and $\sigma_{+}\sigma_{-}\geq1.$ Without
squeezing we obtain the vacuum noise level for two beams $(\Phi_{0}^{(2)})$
where $\sigma(x_{1}\pm x_{2})=\sigma(p_{1}\pm p_{2})=2.$ We observe that
$x_{1}$ and $x_{2}$ are correlated while $p_{1}$ and $p_{2}$ are
anti-correlated both to a level below the vacuum noise level.
This is the same kind of quantum correlations first recorded for the light
from a nondegenerate OPO \cite{Ou92a,Ou92b}. While noise reduction below the
vacuum level is achieved when measuring correlations between the two EPR
beams, the noise from only one of the EPR beams is phase independent and above
the vacuum level. Indeed we find that $V(x_{1,2})=V(p_{1,2})=(\sigma
_{+}+\sigma_{-})/2\geq1,$ where unity is the vacuum level for a single beam
$(\Phi_{0}^{(1)}).$

\begin{figure}[h]
\begin{center}
\includegraphics[width=8.cm]{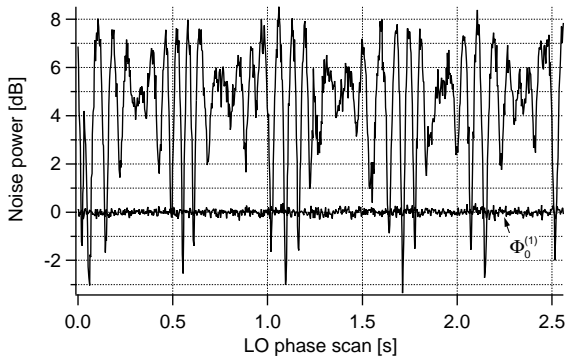}
\end{center}
\caption{Noise of one beam from the EPR beamsplitter as obtained by scanning
the mutual phase difference between the two squeezed beams in addition to a
rapid sweep (about 5 times faster) of the LO in the homodyne detector. The
flat curve represents the vacuum state level for a single beam.}
\label{EPR_scan}
\end{figure}

Experimentally the quality of the EPR state was investigated both by measuring
one EPR beam as well as the correlations between the two beams. The noise in a
single beam was measured while scanning the EPR phase $\theta_{EPR}$ between
the two squeezed beams slowly compared to the scan rate of the local
oscillator in the homodyne detector. An example is presented in Fig.\
\ref{EPR_scan}. The rapid sweep of the local oscillator ensured that both
quadratures were measured for each value of $\theta_{EPR}$ and gives rise to
the fast variations in the trace. As explained above, the noise of a single
EPR beam is expected to be phase independent and arises when overlapping the
two squeezed beams with a mutual phase difference of $\theta_{EPR} = \pi/2.$
Hence, this allows identification of this point in Fig.\ \ref{EPR_scan}, and we
observe excess noise in this case about $5 \: \mbox{dB}$ above the vacuum
noise level. Furthermore, when the two squeezed beams are combined in phase
($\theta_{EPR} = 0$), two squeezed beams exit the beamsplitter giving noise
reduction in this case roughly $3 \: \mbox{dB}$ below vacuum noise.

\begin{figure}[h]
\begin{center}
\includegraphics[width=8.cm]{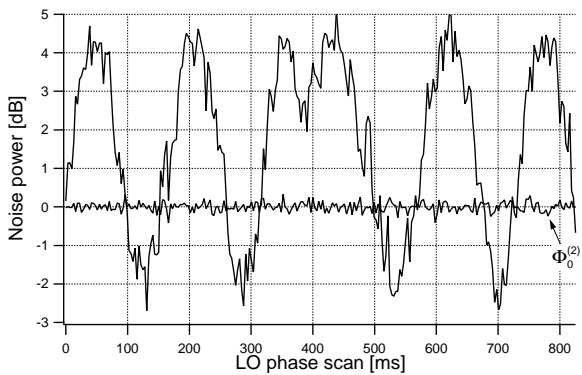}
\end{center}
\caption{Noise obtained by subtracting signals from two homodyne detectors
measuring the two EPR beams while scanning one of the LO phases. The
recorded quantity was $\sigma\left( x_{1}-x_{2}(\theta_{LO})\right) .$ The
flat curve is the vacuum noise level for two beams.}
\label{EPR_2beams}
\end{figure}

A direct measurement of the correlations between the two EPR beams was
obtained via balanced homodyne detection of both EPR beams and subtracting the
resulting photocurrents from the two sets of balanced detectors. In these
measurements the EPR phase was locked at $\theta_{EPR}=\pi/2$ and one of the
homodyne detectors was locked to measure a fixed quadrature. The trace in Fig.\
\ref{EPR_2beams} was recorded by scanning the local oscillator $\theta_{LO}$
of the second homodyne detector, thus recording the variance $\sigma\left(
x_{1}-x_{2}(\theta_{LO})\right)  .$ Reduction below the vacuum level for two
beams $(\Phi_{0}^{(2)})$ was observed when the second homodyne detector
measured the same quadrature as the first homodyne detector, i.e. for
$\theta_{LO}=0.$ We observe correlations of the amplitude quadratures of about
$2\:\mbox{dB}$ with respect to the vacuum level. However, these data were taken
in a non-optimized situation (e.g., inefficient OPO cavity alignment);
the measured degree of squeezing at that time was under $-3\:\mbox{dB}$.
In the actual teleportation experiment inter-beam EPR
correlations of more than $3\:\mbox{dB}$ was obtained.

\section{Quantum teleportation}
\label{QT_results}
Given the preceding description of how to generate the EPR
state, we will now move on to discuss the complete quantum teleportation
protocol. First the electronic servo locks implemented in the experiment are
described, then the technique to calibrate the channels used by Alice to send
classical information to Bob is presented. This establishes the
basis for presenting the experimental results on quantum teleportation.

\subsection{Phase locking}
\label{phaselocking}

\begin{table*}
\begin{ruledtabular}
\begin{tabular}{cll}
NO. & BRIEF DESCRIPTION & LOCKING TECHNIQUE \\ \hline
1 & Doubling cavity resonant to $866\:\mbox{nm}$
  & Pound-Drever-Hall via reflection, $26 \: \mbox{MHz}$ \\
2 & OPO cavity resonant to $866\:\mbox{nm}$
  & Pound-Drever-Hall via transmission, $5\:\mbox{MHz}$ \\
3 & Mode cleaning cavity resonant to $866\:\mbox{nm}$
  & Pound-Drever-Hall via reflection, $26\:\mbox{MHz}$ \\
4, 5 & Relative phase between blue pump ($433\:\mbox{nm}$) and injected
       & Lock-in via $1\%$ pick-off of cavity transmission \vspace{-3pt} \\
  & \hspace{0.2in} beams 1 \& 2 ($866\:\mbox{nm}$) to zero (maximum gain) & \\
6 & Relative phase between two squeezed beams to $\pi/2$ (to
  & DC interference fringe \vspace{-3pt} \\
  & \hspace{0.2in} create EPR state) & \\
7, 8 & Alice's LOs to $x$ and $p$ quadratures & RF interference fringe,
     $3\:\mbox{MHz}\rightarrow x, 5\:\mbox{MHz}\rightarrow p$ \\
9 & Bob's coherent beam to $x$ quadrature
  & RF interference fringe, $3\:\mbox{MHz}$ \\
10 & Victor's LO to either $x$ or $p$ quadratures & RF interference fringe,
     $3\:\mbox{MHz}\rightarrow x, 5\:\mbox{MHz}\rightarrow p$ \\
\end{tabular}
\end{ruledtabular}
\caption{Summary of the servo systems that were implemented. See main text for
  further discussion.}
\label{tab_locking}
\end{table*}

In the complete quantum teleportation experiment ten servo systems in total
were implemented for locking optical phases and cavities. A summary of the
locking techniques is given in Table \ref{tab_locking}. Three of the servos
(numbers $1-3$) are for locking cavities and use the standard Pound-Drever-Hall
technique where reflection or transmission of RF modulation sidebands are used
to derive an error signal. The frequency doubling cavity was locked by
observing the cavity reflection dip off the input coupler, and generating an
error signal using RF sidebands at $26 \: \mbox{MHz}$. The OPO cavity was
locked by observing a $1\%$ pick-off from the transmitted light from a weak
coherent beam injected through a high reflector. The injected signal carried
$5 \: \mbox{MHz}$ sidebands for locking. Finally, the mode-cleaning cavity was
locked in reflection using sidebands at $26 \: \mbox{MHz}$.

The remaining seven servos were used to keep the optical phases properly
aligned for successful teleportation \cite{vanenk01}. Two weak injection beams
were seeded into the two counterpropagating modes of the OPO and used for
several purposes: to align the OPO cavity and the homodyne detectors, and to
use for locking of optical phases. The two injected signals were phase
modulated at $3\:\mbox{MHz}$ (injection 1) and $5\:\mbox{MHz}$ (injection 2)
respectively, using an electro-optical modulator (EOM) in each signal path.
The relative phase differences between the pump beams and the injected beams
were locked using standard lock-in techniques where the error signal is derived
from the phase sensitive amplification (due to the parametric gain in the OPO)
of the injected signal. The phases are locked at maximum gain corresponding to
the situation where the direction of the phasor of the coherent injected field
is along the long axis of the squeezing ellipse of the squeezed vacuum beams.

The third phase-lock, the EPR phase lock, keeps the two independent squeezed
beams incident on the EPR beamsplitter with a phase difference of $\pi/2$ to
ensure the production of the EPR state. This was done by using $1\%$ leakages
from mirrors in the EPR beam paths. The DC interference signal between
the two injected beams from the two distinct EPR paths was subtracted to
produce an error signal centered around zero. This zero crossing corresponds
to $\pi/2$ phase difference between the two squeezed beams.

To lock the local oscillators in Alice's two homodyne detectors, the RF beat
notes between the LO and the $3$ and $5 \: \mbox{MHz}$ sidebands on the
injected beams were demodulated to produce the respective error signals for
Alice I $(3 \: \mbox{MHz})$ and II $(5 \: \mbox{MHz})$. As the EPR phase lock
described above already keeps the two injected beams at $\pi/2$ phase
difference, the LOs in Alice I and II will also stay at $\pi/2$ relative phase
difference. Bob's LO phase was locked using the RF interference fringe at $3
\: \mbox{MHz}$ between itself and the modulation sidebands transported with
EPR2. This means that Alice I is locked to the same quadrature as Bob, which,
by an arbitrary convention, sets Alice I as $x$ and Alice II as $p$.
Thus to complete the classical information channel, Alice I's measured output
(photocurrent) was sent to Bob's amplitude modulator, and Alice II's output was
sent to Bob's phase modulator. The teleported state then emerges from
Bob's beamsplitter, and is sent to Victor for verification. Victor's LO phase
can be either scanned, or locked to either $3$ or $5 \: \mbox{MHz}$ to check
$x$ and $p$ separately.

\subsection{Classical information channels}
\label{class_info_chans}

A crucial part of the teleportation protocol is the transmission of classical
information from Alice to Bob. In the present experiment the classical
information is just the photocurrents from Alice's two homodyne detectors.
These signals have to be faithfully transmitted to Bob without distortion and
with proper phase and gain, and for that reason several RF amplifiers, filters
and delay boxes were used in the classical channel paths. We typically
measured Alice's and Victor's noise levels at $1.475 \: \mbox{MHz}$, thus the
electronics of the channels were optimized at that frequency. In the following
we will present a method to perform the calibration of these classical channels.

To ensure that we are operating at a gain $g_x=g_p=1$ such that $\beta_{V} =
\beta_{in}$, we compared the photocurrents measured by Alice and Victor when
there were no EPR beams present for the case of a coherent state of amplitude
$\beta_{in}$ sent to Alice as the input state. In practice, this was easily
achieved by blocking the optical beam paths of the EPR state. In this case,
it can be shown that when an amplitude or phase modulated beam is sent to
Alice as the input state, the ratio between the spectral densities measured
at Victor and Alice is given by
\begin{equation}
\frac{\Phi_{V}}{\Phi_{A}} = \frac{2}{\xi_2^2\eta_{A}^2},
\end{equation}
where we have assumed Alice and Victor are measuring the same quadrature
(either $x$ or $p$), that $\left\vert\beta_{in}\right\vert^2 \gg 1$ and that
the efficiencies are close to unity. We observe that Victor records a spectral
density (i.e., noise power of the RMS photocurrent) two times (corresponding
to $3\:\mbox{dB}$) higher than Alice $x$ (or $p$). This factor of two can
easily be understood since the input beam is split into two equal halves at
Alice's 50/50 beamsplitter. Hence, this identifies a signature for the optimum
condition of the classical gain.

On the other hand, for vacuum input, the ratio between Victor's and Alice's
spectral densities, or equivalently, their variances since now
$\left\vert\beta_{in}\right\vert^2 = 0$, is found to be
\begin{equation}
\label{victor_var}
\frac{\sigma_{V}}{\sigma_{A}} = 1 + \frac{2}{\xi_2^2\eta_{A}^2},
\end{equation}
when the classical gain is optimum. This means Victor's output is $\approx 3$
times higher than Alice $x$ (or $p$) again in the limit where all detector
efficiencies are close to unity.

\begin{figure}[htb]
\begin{center}
\includegraphics[width=8.cm]{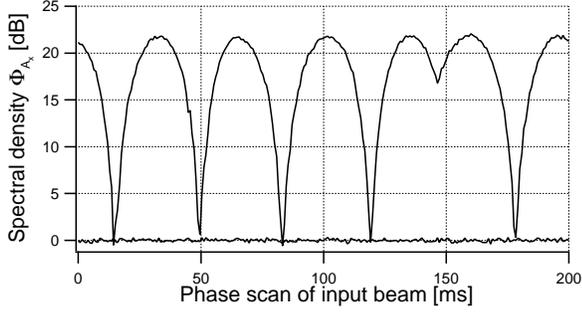}
\end{center}
\caption{Spectral density $\Phi_{Ax}(\Omega)$ relative to the vacuum level
recorded by Alice $x$ while scanning the phase of the input beam
both for vacuum (flat trace), and an input beam with amplitude modulation at
$24.9\:\mbox{dB}$ above the vacuum noise level. The measured amplitude of this
beam is $21.9\:\mbox{dB}$ or $3\:\mbox{dB}$ lower than the actual input, as
explained in the text. The measurement frequency was $1.475\:\mbox{MHz}$, RBW
$30\:\mbox{kHz}$, and VBW $1\:\mbox{kHz}$.}
\label{Alice1_unknown}
\end{figure}

\begin{figure}[htb]
\begin{center}
\includegraphics[width=8.cm]{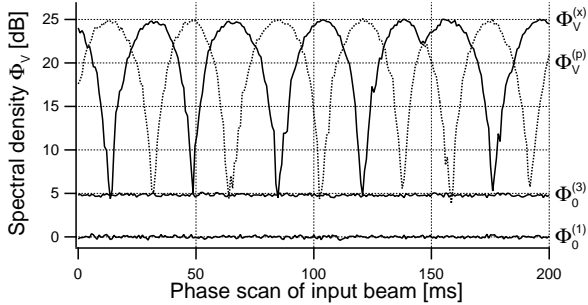}
\caption{Spectral density $\Phi_{V}(\Omega)$ relative to the vacuum level
recorded by Victor while scanning the phase of the input beam.
$\Phi_{0}^{(1)}$ is Victor's noise level for one unit of vacuum, while
$\Phi_{0}^{(3)}$ shows the 2 extra units making 3 total units of vacuum noise
measured in an attempt to recreate Alice's vacuum state input without any
entanglement. $\Phi_{V}^{(x)}$ and $\Phi_{V}^{(p)}$ show the recreation of
Alice's input coherent modulation amplitude of $24.9\:\mbox{dB}$,
demonstrating that the peak input and output amplitudes are equal, and the
$x$ and $p$ quadratures are indeed $\pi/2$ apart in phase. Victor's LO was
phase-locked to the $x$ and $p$ quadratures respectively for these
measurements.}
\label{Vic_unknown_clas}
\end{center}
\end{figure}

Fig.\ \ref{Alice1_unknown} shows the spectral density $\Phi_{Ax}(\Omega)$
of photocurrent fluctuations recorded at Alice $x$ from input beams with
modulation amplitudes corresponding to $24.9\:\mbox{dB}$ and $0\:\mbox{dB}$
(vacuum), respectively. The signal recorded at Alice $p$ mirrors this trace,
except that the coherent amplitude is shifted in phase by $\pi/2$,
demonstrating that Alice $x$ and $p$ are $\pi/2$ apart in phase, as required.
The corresponding traces for the spectral density $\Phi_{V}(\Omega)$\ for
Victor's detector are shown in Fig.\ \ref{Vic_unknown_clas}. Here we show
explicitly the $\pi/2$ phase shift when Victor's LO is phase-locked to the
$x$ and $p$ quadratures, respectively. We see that Victor records
$3\:\mbox{dB}$ higher spectral density for the amplitude modulated input
and $4.8\:\mbox{dB}$ greater for the vacuum input, which indicates that
the gain of the classical channels has been properly calibrated relative to
the criteria of Eqs. (\ref{fidelity_exp}) and (\ref{sigma_x}).

We now turn our attention to the phases of the RF signals, keeping in mind the
distinction between the phase of the optical carrier, and the phase of the
RF signal at our measurement frequency, $1.475\:\mbox{MHz}$. We have already
discussed optical phases in detail in Sec.\ \ref{phaselocking} above, and
indeed implicitly assumed correct optical phases in the discussion on
finding optimal gain. In discussing RF phase, we will continue to assume that
our optical phase-lock servos are working correctly.

Our goal is quantum noise subtraction at our analysis frequency, between the
correlated beams EPR1 (measured at Alice and sent to Bob via the classical
channels) and EPR2 at Bob's beamsplitter. Obviously, the best way to achieve
this goal is for the relative RF phases of EPR1 and EPR2 to be zero for
perfect subtraction. However, this is quite impractical in the current
laboratory setup since EPR2 arrives at Bob's beamsplitter directly from the
EPR beamsplitter, while EPR1 takes a very indirect route involving electrical
photocurrents that travel much slower than light. It is sufficient, therefore,
to ensure that the relative phase difference is a factor of $2\pi$ by
implementing delays in the classical channels between Alice and Bob. In
practice, we can also keep the phase difference any multiple of $\pi$ and
compensate for this by flipping the sign of Bob's optical phase-lock error
signal. In this way, Bob \textit{adds} EPR1 and EPR2 instead of subtracting
them, thereby optically creating a $\pi$ RF phase shift that compensates for
the $\pi$ phase delay in the classical channels.

Finally, Bob's two modulators must provide pure phase and amplitude modulation,
respectively. This condition is satisfied by carefully controlling the input
beam polarizations for the two temperature compensated EOMs.

Some care must be given to the maximum time delay allowed in our classical
channels. This is set by the OPO linewidth (HWHM) of $5.4\:\mbox{MHz}$,
corresponding to a correlation time between EPR1 and EPR2 of about
$30\:\mbox{ns}$ if the full bandwidth of the OPO were employed for
teleportation. For a more detailed discussion see Ref.\ \cite{vanLoock00}.

However, in our experiment, a much smaller effective bandwidth is employed
corresponding to the detection bandwidth for Alice, the bandwidth of the
classical channel from Alice to Bob, and the frequency range of Bob's
modulators. Finally, there is the bandwidth employed by Victor in his
verification of the protocol. For simplicity, here we assume that the
effective detection bandwidth of our protocol is equal to the RF bandwidth,
$\nu$, of the spectrum analyzer employed by Victor in his analysis,
typically around $\nu=30\:\mbox{kHz}$ (see figures in Sec.\
\ref{tptresults}). The relevant issue is the ratio between our
analysis frequency, $\Omega /(2\pi)=1.475\:\mbox{MHz}$, and $\nu$.
We see that this ratio is about $50$ cycles. Thus $2\pi$ of RF phase
delay in our classical channels contributes to a roughly $2\%$
effect on the noise subtraction quality, which is small but not negligible.

When the gain and RF phase of the classical channels as well as the optical
phases at Alice's and Bob's detectors were suitably optimized, Victor
recorded a stable output while the phase of the unknown input state was being
scanned, independent of the input state amplitude and phase over a wide range,
as discussed below. In the case of vacuum input we obtain the flat trace at
$4.8\:\mbox{dB}$ (trace $\Phi_{0}^{(3)}$) in Fig.\ \ref{Vic_unknown_clas}.
The trace remained stable for tens of minutes, with fluctuations on the order
of $\pm 0.1\:\mbox{dB}$.

It was not a trivial task to realize the balance of the two classical channels
due to above-mentioned reasons. One practical way that we employed to optimize
the system and to judge the effectiveness of the two classical channels was to
send RF modulated optical fields at $1.475\:\mbox{MHz}$ through the two
injection ports of the OPO cavity. These modulated optical signals were
allowed to propagate to Alice and Bob, just as the EPR beams would in the
presence of blue pump. We could then easily optimize the subtraction of the
two classical fields at Bob's 99/1 beamsplitter. In terms of the
conditions stated above, this \textit{classical noise} subtraction directly
mimics the \textit{quantum noise} subtraction that we perform using the
entangled EPR state during quantum teleportation, which means that if we
obtain good classical subtraction, we should in fact be operating at the
optimal conditions for quantum teleportation. Typically the subtraction was
about $25\:\mbox{dB}$ for each channel.

\begin{figure}[htb]
\begin{center}
\includegraphics[width=8.cm]{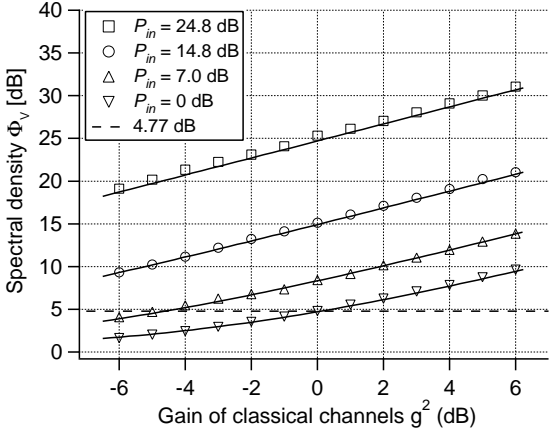}
\end{center}
\caption{Measured spectral density $\Phi_{V}(\Omega)$ from Victor's
balanced homodyne detector
as a function of gain for different amplitude-modulated coherent
state and comparison with theory. The parameters for the theoretical curves
were the measured visibilities: $\xi_{1}=0.985$, $\xi_{2}=\xi_{3}=0.994$,
$\xi_{4}=0.985$, $\xi_{5}=0.985$ and the detector quantum efficiency was
$\alpha=0.988$. The measurement frequency was $1.475\:\mbox{MHz}$. $P_{in}$
corresponds to the actual power of the input state presented to Alice,
in dB above the vacuum state.}
\label{victor_change_gain}
\end{figure}

In order to check the linearity and dynamical range for the classical
channels, we measured Victor's output noise levels as a function of gain for
input modulation sidebands of various amplitudes when the two classical
channels were balanced. Fig.\ \ref{victor_change_gain} shows the results for
input beams with the following modulation amplitudes: $0\:\mbox{dB}$ (vacuum),
$7.0\:\mbox{dB}$, $14.8\:\mbox{dB}$ and $24.8\:\mbox{dB}$, where in each case
the full $2\pi$ of phase variation was explored. It shows that the linearity
is very good from $0\:\mbox{dB}$ to $25\:\mbox{dB}$ input modulation
amplitudes, which means we can teleport any coherent state amplitude within
that range. The theoretical traces on the figure are based on our measured
efficiencies without any adjustable parameters.

\subsection{Teleportation results}
\label{tptresults}
\begin{figure}[htb]
\begin{center}
\includegraphics[width=8.cm]{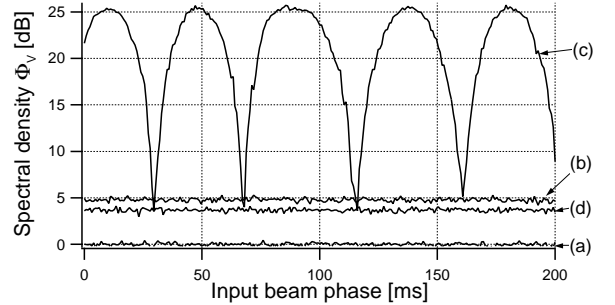}
\end{center}
\caption{Spectral density $\Phi_{V}(\Omega)$
recorded by Victor with the phase of the input beam scanning. Trace (a)
is Victor's shot-noise level, trace (b) marks the 3 units of vacuum noise
measured without entanglement, trace (c) is one quadrature of the teleported
output coherent state, and trace (d) is the teleported output vacuum state
(see text for details). For trace (c) Victor's LO was phase-locked to the $x$
quadrature, while for traces (a), (b) and (d) Victor's LO was freely scanned,
and a ten trace average was used. The average OPO pump power was
$34\:\mbox{mW}$ per beam, measurement frequency $1.475\:\mbox{MHz}$,
RBW $30\:\mbox{kHz}$, and VBW $1\:\mbox{kHz}.$}
\label{Vic_unknown}
\end{figure}

Fig.\ \ref{Vic_unknown} shows quantum teleportation results for a coherent
input state. All traces are Victor's measured variances at
$1.475\:\mbox{MHz}$ and at pump power of $33/35\:\mbox{mW}$ in the OPO paths
1 and 2, respectively. The amplitude of the input state was about
$25.5\:\mbox{dB}$ higher than the vacuum level. Trace (a) is one unit of the
vacuum noise, that is, the vacuum-state level or shot-noise level (SNL) of
Victor, which is obtained by blocking Bob's beam. Trace (b) marks the 3
units of vacuum noise in the case of absent EPR beams but with Alice and
Bob engaged nonetheless in the teleportation protocol, which is
$4.8\:\mbox{dB}$ above the SNL with our efficiencies close to 1 and is
obtained by blocking the blue pumps in the experiment. Trace (c) shows the
phase sensitive noise when the EPR beams and the AM sidebands on the input
state are present, while Victor's LO is phase-locked to the $x$ quadrature.
Locking Victor's LO to the $p$ quadrature produces an analogous trace with
the peaks offset by $\pi/2$ in phase. Closer inspection of traces such as
(c) in Fig.\ \ref{Vic_unknown} shows that the minimum noise level is
approximately $1.1\:\mbox{dB}$ below the level of three units of the vacuum,
although it is rather difficult to get an accurate reading because of the
mismatch of scan rate and detection bandwidths. This noise level corresponds
to $2.3$ vacuum units. The peak of the trace should have the same amount of
noise reduction, that is, from $354.8$ to $354.1$ vacuum units
(from $25.50\:\mbox{dB}$ to $25.49\:\mbox{dB}$), but this reduction is too
small to observe in the graph. Trace (d) corresponds to the vacuum input
state, which is obtained by blocking the modulated input beam. Acquisition
parameters are: resolution bandwidth $30\:\mbox{kHz}$, video bandwidth
$1\:\mbox{kHz}$ and sweep time $200\:\mbox{ms}$.

\begin{figure}[htb]
\begin{center}
\includegraphics[width=8.cm]{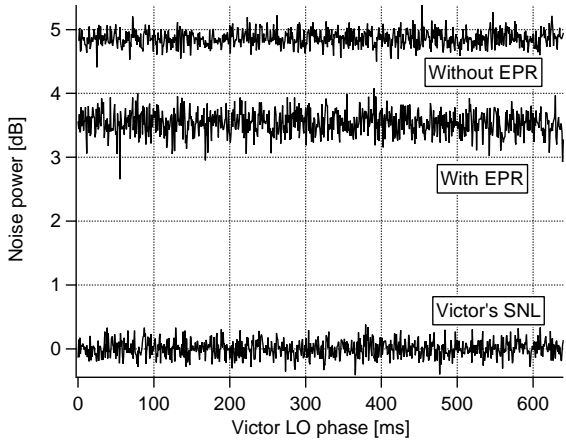}
\end{center}
\caption{Noise recorded by Victor showing in detail the reduction in the noise
level with EPR beams (entanglement) present. With EPR beams, the measured
variance at Victor was $\sigma_{V}^{x}=\sigma_{V}^{p}=3.54\pm 0.19\:\mbox{dB}$.
The average OPO pump power was $40\:\mbox{mW}$ per beam, and acquisition
parameters are the same as in Fig.\ \ref{Vic_unknown}. All traces use a ten
trace average. SNL stands for shot-noise level.}
\label{QT_trace}
\end{figure}

The best noise reduction that we have obtained to date is shown in detail
in Fig.\ \ref{QT_trace}. With the EPR beams present, the variances recorded by
Victor are $\sigma_{V}^{x}=\sigma_{V}^{p}=3.54\pm 0.19\:\mbox{dB}$, while
with the EPR beams absent, $\sigma_{V}^{x}=\sigma_{V}^{p}=4.86\pm
0.12\:\mbox{dB}$. The entanglement of the EPR beams thus leads to a quantum
noise reduction of $1.32\pm 0.16\:\mbox{dB}$. This result was obtained
with $40\:\mbox{mW}$ pump power in each OPO path. The measurement parameters
are the same as that in Fig.\ \ref{Vic_unknown} except that the sweep time
was $640\:\mbox{ms}$ and we use a ten trace average for all traces. For this
particular trace, the measured detection efficiencies were characterized by
$\xi_{1}=0.986, \xi_{2}=\xi_{3}=0.990, \xi_{4}=0.980, \xi_{5}=0.975$, and
$\alpha_{Ax}=\alpha_{Ap}=\alpha_{V}=0.988$.

\begin{figure}[htb]
\begin{center}
\includegraphics[width=8.cm]{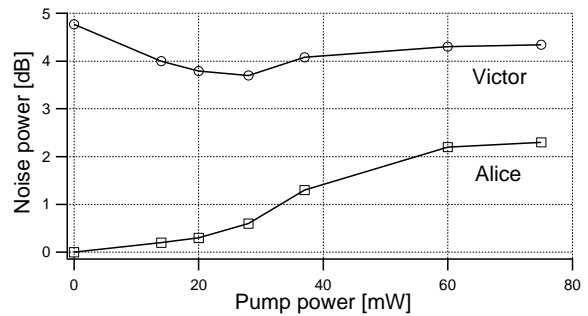}
\end{center}
\caption{Noise measured by Alice and Victor as functions of the pump of the
OPO. The data points have been connected to ease viewing.}
\label{Vic_pump}
\end{figure}

A study of the dependence of the output variance on OPO pump power yields
further insight into the experiment. Fig.\ \ref{Vic_pump} shows the variances
of Alice and Victor as functions of pump power. From the squeezing results
in Fig.\ \ref{squeez_pump}, we can see that as the pump increases, both the
squeezing and anti-squeezing increase, even though the squeezing increases
very slowly. Therefore, the entanglement becomes stronger with increasing
pump power. This phenomenon is reflected in the data shown in Fig.\
\ref{Vic_pump} at pump powers below $30\:\mbox{mW}$, where Alice's variance
increases with pump power, whereas Victor's variance drops below
$4.8\:\mbox{dB}$, as predicted. The best noise reduction for Victor was
around $30\:\mbox{mW}$ of blue pump for this particular set of data. Higher
pump power did not help to reduce the noise; it instead increased both
Alice's and Victor's variances even though we expect Victor's variance to
continue to decrease, or at least remain stable. The likely culprits
responsible for this degradation in performance will be discussed in the next
sections. Chief among them is the performance of the various locking servos.
As discussed in Sec.\ \ref{phase_qt}, fluctuations around the nominal ideal
settings (e.g., $\pi/2$ for the squeezed beams that form the EPR beams)
allows excess noise to contaminate the ``quiet'' quadratures, an effect that
becomes more important as the degree of squeezing is increased.

\subsection{Teleportation fidelity}
\label{infer_F}

From the measured variances reported in Sec.\ \ref{tptresults} above, we can
gauge the quality of the protocol by inferring the fidelity $F$ via Eqs.\
(\ref{fidelity_exp}) and (\ref{sigma_x}). We first assume that the noise in
the teleported output state $\rho_{out}$ obeys Gaussian statistics
\cite{Furusawa98,Braunstein00}. From the data in Fig.\ \ref{QT_trace}, where
$\sigma_{V}^{x}=\sigma_{V}^{p}=3.54\pm 0.19\:\mbox{dB}$, and where the gain
$g$ has already been set to unity by the techniques described in Sec.\
\ref{class_info_chans}, direct application of Eq.\ (\ref{fidelity_exp})
gives $F=0.61\pm 0.02$.

Note that no correction whatsoever has been applied to this result; it
corresponds to the fidelity obtained directly from Victor's photocurrent.
We can certainly attempt to infer the fidelity associated with the field
emerging from Bob's beamsplitter, rather than the photocurrent detected by
Victor. To do so, we return to the issue pointed out earlier that we
calibrate $g_{x,p}$ by ensuring $\beta_{V} = \beta_{in}$ as opposed to
$\beta_{out} = \beta_{in}$. From the discussion in Sec.\ \ref{Fidelity_vis},
we see that if Victor had unit detection efficiency $(\xi_5=\eta_V=1)$,
this issue would not arise at all. In our experiment where
$(\xi_5<1,\eta_V<1)$, the actual unnormalized gains $g_{(0)}$ were set to
be too large by a factor $\left(\xi_5\eta_V\right)^{-1}$ than that
neccessary for optimal reconstruction of Bob's field, instead of Victor's
photocurrent. We can use Eq.\ (\ref{sigma_x}) to compare Victor's variances
when $(\xi_5=\eta_V=1)$ and when $(\xi_5<1,\eta_V<1)$ with the same degree of
squeezing as in Fig.\ \ref{QT_trace}. We thus infer that if Victor had
perfect detectors, the variance of Bob's teleported output field (or
equivalently now, Victor's variance as measured by his photocurrent) is
given by $\sigma_{W}^{(x,p)}=\sigma_{V}^{(x,p)}=3.47\:\mbox{dB}$ above the
shot-noise level, which corresponds to an inferred fidelity of $F_B=0.62$.

Returning to fidelity referenced to Victor's photocurrent, we estimate that
with $42\:\mbox{mW}$ pump power in each OPO path, where we have measured
$-3.73\:\mbox{dB}$ squeezing and $6.9\:\mbox{dB}$ anti-squeezing at Victor
(see Fig.\ \ref{squeez_trace}), the EPR entanglement at the EPR beamsplitter
is characterized by the factors: $\sigma^{-}=-3.97\:\mbox{dB}$ and
$\sigma^{+}=7.0\:\mbox{dB}$. These numbers were obtained by back-propagating
the squeezed beams to the EPR beamsplitter from Victor's homodyne detectors,
considering the effects of Victor's homodyne efficiency $\xi_{5}=0.972$,
photodiode quantum efficiency $\alpha_V=0.988$ and the EPR homodyne
efficiency $\xi_{EPR}=0.985$. With measured efficiencies $\xi_{1}=0.985$,
$\xi_{2}=\xi_{3}=0.990$, $\xi_{4}=0.980$, $\xi_{5}=0.975$, the predicted
variance of the teleported output state emerging from Bob's beamsplitter is
$\sigma_{W}^{(x)}=\sigma_{W}^{(p)}=2.82\:\mbox{dB}$. The inferred fidelity
would then be $F_P=0.69$.

By contrast, our best entanglement-assisted noise reduction measured by
Victor's balanced detector is $1.32\:\mbox{dB}$ below the level with no
EPR beams, which corresponds to $\sigma_{V}^{(x)}=\sigma_{V}^{(p)}=
3.54\:\mbox{dB}$ and an inferred fidelity $F=0.61$, as has been discussed.

\begin{figure}[htb]
\begin{center}
\includegraphics[width=8.cm]{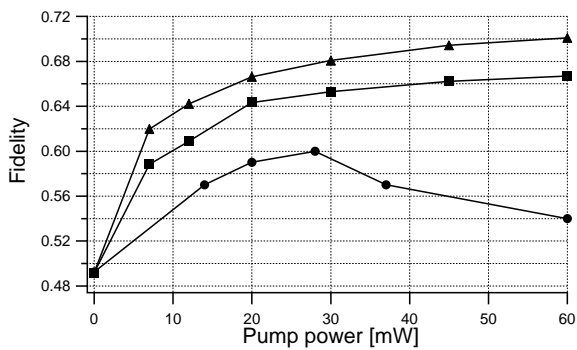}
\end{center}
\caption{Fidelity as a function of the OPO pump. Dots are the experimentally
measured fidelities, squares are the expected fidelities given the measured
degrees of squeezing shown in Fig.\ \ref{squeez_pump}, and triangles are the
expected fidelities based on our theoretically predicted degrees of squeezing
in Fig.\ \ref{squeez_pump}. Again, the lines connecting the data points are
to ease viewing.}
\label{Fidelity_pump}
\end{figure}

To gain more insight into this discrepancy, we performed an analogous study
to that shown in Fig.\ \ref{Vic_pump}, where we plot the inferred fidelities
from the measured variances versus pump power in Fig.\ \ref{Fidelity_pump}.
The inferred experimental fidelities peak at around $30\:\mbox{mW}$
corresponding to the minimum in Victor's measured variance. Here we also
show the values of $F$ that we might be able to reach with the current
apparatus. First of all, the square symbols in Fig.\ \ref{Fidelity_pump}
derive from the inferred degrees of squeezing at the EPR beamsplitter that
are deduced from the data in Fig.\ \ref{squeez_pump}. This was done from the
measurement results by back-propagating the squeezed beams from Victor's
homodyne detector to the EPR beamsplitter as previously described. In
addition, the triangles in Fig.\ \ref{Fidelity_pump} correspond to the
inferred fidelities from the theoretically predicted degrees of squeezing
given by the solid line in Fig.\ \ref{squeez_pump}. It can be seen that the
disagreement between the predicted and measured fidelities is already apparent
at low pump powers. However, the mismatch becomes more pronounced at higher
pump powers where the measured fidelities start to decrease rather than
increase with the pump level.

There are several factors contributing to the discrepancy between the measured
and predicted variances and fidelities. The first is phase fluctuations from
different locking systems, especially the EPR phase-lock. As discussed in
Sec.\ \ref{phase_qt}, phase fluctuations add extra noise to the quadratures,
and this effect is more pronounced at higher pump powers where the
anti-squeezed quadrature is large. Another factor is the bandwidth of our two
classical channels, which by bad design on our part, has been found to have
excessive phase variation over the relevant RF bandwidth for the undistorted
transmission of classical information. In fact, by an aforementioned
technique, we measured the actual subtraction of the RF signal at
$1.475\:\mbox{MHz}$ through the classical channels and EPR2. It was found
that frequency offsets of about $5\:\mbox{kHz}$ led to drops in the
cancellation from $-25\:\mbox{dB}$ to $-20\:\mbox{dB}$. When the offset was
$20\:\mbox{kHz}$, the cancellation is only $-9\:\mbox{dB}$. Increasing the
bandwidth while keeping relatively high isolation for the filter for various
control signals (e.g., modulations at $3$ and $5\:\mbox{MHz}$) in the
classical channels will be helpful. A third reason for the discrepancy in
Victor's variances is the imperfect character of Bob's EOMs, which cause
coupling between the $x$ and $p$ quadratures, again resulting in
contamination between squeezing and anti-squeezing. In order to reach higher
fidelity, we are working to improve these aspects of the experiment.

\section{Conclusion}
\label{conclusion}
We have described the details of recent experimental work to perform quantum
teleportation for continuous variables. We have discussed a real experimental
system where we considered many of the prevalent loss sources in our
experiment, thus providing a detailed analysis of how the variances measured
at Alice and Victor during teleportation vary with squeezing and ultimately
OPO pump power. Phase fluctuations due to imperfect locking systems were also
discussed, and it has been shown that nonideal detection schemes as
well as phase fluctuations eventually degrade the noise reduction
recorded by Victor and consequently reduce the teleportation fidelity.

We have discussed how to prepare experimentally the entangled EPR beams.
Our entangled EPR fields at the EPR beamsplitter were typically characterized
by $\sigma^{-}\simeq-4\:\mbox{dB}$ and $\sigma^{+}\simeq7\:\mbox{dB}$
according to our measured squeezing and efficiencies, which implies that our
measured prospective fidelity would be $0.69$.

The experimental setup and procedure was described in detail, including
the OPO, the source of the entangled EPR fields. We have discussed optical
phase-lock servo systems that ensure Alice is able to correctly measure the
two orthogonal quadratures, $x$ and $p$. Lastly, we have also described our
classical channels through which classical information in the form
of photocurrents obtained by Alice can be sent to Bob with the goals of
minimal distortion and proper phase and gain in order to be sure that Bob
can use that information to recover the original input state.
The teleportation procedure was investigated for arbitrary unknown
coherent states with any phase and a wide range of amplitudes.

The experiment clearly revealed the quduties that limit the classical
performance of such a teleportation system. Victor unavoidably measures two
extra units of vacuum noise if there is no entanglement. By employing the
entangled fields of the EPR state, we demonstrated that the quduties were
suppressed by $-1.32\pm 0.16\: \mbox{dB}$, which corresponds to an inferred
fidelity of $F=0.61\pm 0.02$ for coherent states, which when corrected for
the efficiency of detection by Victor, gives a fidelity of $0.62$. The
apparatus was shown to succeed for arbitrary coherent states with amplitudes
up to $25 \: \mbox{dB}$ above the vacuum level. This demonstrates that with
entanglement, the procedure exhibits better performance than the classical
bound of fidelity $F=0.50$ \cite{Braunstein00,Braunstein01}, and hence is
genuinely a quantum protocol for unconditional teleportation.

We discussed Alice's and Victor's measured variances as functions of
OPO pump power. The data in Fig.\ \ref{Vic_pump} showed that at low pump
powers, Alice's measured variance $\sigma_A$ increased and Victor's measured
variance $\sigma_V$ decreased with pump power, as expected. However, at
higher pump powers (above $\approx 30\:\mbox{mW}$), while $\sigma_A$ continued
to increase with pump as expected, $\sigma_V$ did not continue to decrease,
but began increasing instead. This implied that the
anti-squeezing quadrature contaminates the squeezed quadrature at high pump
powers. Possible reasons for this contamination include
fluctuations in the phase-lock servos, limited classical channel
bandwidth and impure amplitude and phase modulators at Bob's station.

It is encouraging to note that our high detection efficiencies together with
the relatively high degree of entanglement that we have achieved shows that
our apparatus is capable of producing higher fidelity between the input and
output states. In addition, a new scheme with the OPO pumped only by a single
unidirectional blue beam to form the EPR state is being planned. We could
then hope to obtain over $-5 \: \mbox{dB}$ of entanglement by mitigating
high BLIIRA, and thus reducing the intracavity losses in the OPO. Such
capabilities would be of interest to quantum information processing with
continuous quantum variables \cite{BraunsteinPati00}.

\begin{acknowledgments}
The contributions of J.\ Buck and A.\ Furusawa are gratefully acknowledged.
This research was funded by the National Science Foundation, by the Office
of Naval Research, by the Caltech MURI on Quantum Networks administered
by the Office of Army Research, and by the NSF sponsored Institute for
Quantum Information. TCZ was partially supported by the EYTP of MOE, P.~R.~C.
\end{acknowledgments}


\begin{thebibliography}{}
\bibitem{ANU02}
Recent press reports have described continuous variable teleportation from
the group of H. Bachor and P.~K. Lam at the Australian National University.
Unfortunately, there is no scientific manuscript on which to base an
assessment of these claims.
\end{thebibliography}
\end{document}